\documentclass[aps,twocolumn,preprintnumbers,floats]{revtex4}
\usepackage[colorlinks, citecolor=blue,anchorcolor=red,menucolor=red, linkcolor=red,filecolor=red,urlcolor=blue,frenchlinks=red]{hyperref}
\usepackage{amsfonts}
\usepackage{amsmath}
\usepackage{amssymb}
\usepackage{CJKutf8}
\usepackage{color}
\usepackage{comment}
\usepackage{epsfig}
\usepackage{epstopdf}
\usepackage{float}
\usepackage{graphicx,booktabs}
\usepackage{indentfirst}
\usepackage{longtable,lscape}
\usepackage{mathrsfs}
\usepackage{mathtools}
\usepackage{morefloats}
\usepackage{pifont}
\usepackage{txfonts}
\begin{document}

\begin{CJK}{UTF8}{gbsn}

\title{Hidden and double charm-strange tetraquarks and their decays in a potential quark model}
\author{Feng-Xiao Liu$^{1,4}$ \footnote {E-mail: liufx@ihep.ac.cn}, Ru-Hui Ni$^{2,4}$, Xian-Hui Zhong$^{2,4}$ \footnote {E-mail: zhongxh@hunnu.edu.cn}, Qiang Zhao$^{1,3}$ \footnote {E-mail: zhaoq@ihep.ac.cn}}

\affiliation{ 1) Institute of High Energy Physics, Chinese Academy of Sciences, Beijing 100049, China}

\affiliation{ 2) Department of Physics, Hunan Normal University, and Key Laboratory of Low-Dimensional Quantum Structures and Quantum Control of Ministry of Education, Changsha 410081, China }

\affiliation{ 3) University of Chinese Academy of Sciences, Beijing 100049, China}

\affiliation{ 4) Synergetic Innovation Center for Quantum Effects and Applications (SICQEA),
Hunan Normal University, Changsha 410081, China}

\begin{abstract}

We carry out a systematic study of the $1S$-wave hidden and double charm-strange tetraquarks $cs\bar{c}\bar{s}$ and $cc\bar{s}\bar{s}$ in a nonrelativistic potential quark model framework with the explicitly correlated Gaussian method, and the mass spectra, color-spin configurations and possible decay modes are obtained. We find that although these states are all above their open flavor thresholds, their rearrangement decay widths are rather narrow which can be understood by the mismatching of the wave functions between the initial and final states. It implies that the tetraquarks of $cs\bar{c}\bar{s}$ and $cc\bar{s}\bar{s}$ may have a good chance to exist as genuine tetraquark states. It also shows that the color-spin configurations of the $cs\bar{c}\bar{s}$ and $cc\bar{s}\bar{s}$ systems are quite different. We find that for a physical state of $cs\bar{c}\bar{s}$ its color configurations can be dominated by either the $|11\rangle_{c}$ or $|88\rangle_{c}$ ones. It suggests that some hidden charm-strange tetraquark states may strongly couple to two color-singlet hadrons if the kinematics and dynamics allow. In contrast, we find that the color configurations $|11\rangle_{c}$ and $|88\rangle_{c}$ in a double charm-strange $cc\bar{s}\bar{s}$ state are rather compatible. It may suggest that an overall color-singlet tetraquark (i.e. a genuine color-singlet) should always play a role in the $T_{cc\bar{s}\bar{s}}$ states. Discussions taking into account some experimental candidates are presented, and suggestions on further experimental searches are also made.

\end{abstract}


\maketitle

\section{Introduction}

Since the $X(3872)$ was first observed at Belle in 2003~\cite{Belle:2003nnu}, many candidates of exotic hadrons have been reported by experiments such as \emph{BABAR}, Belle, LHCb, and BESIII, etc. (e.g. see recent reviews~\cite{Chen:2016qju,Esposito:2016noz,Karliner:2017qhf,Guo:2017jvc,Brambilla:2019esw}). The masses of these exotic states are generally located near particular thresholds of two conventional open-heavy-flavor mesons. Some of these states found in the $J/\psi\phi$ or $D_{s}^{+}D_{s}^{-}$ invariant mass spectrum may contain both a $c\bar{c}$ pair and an $s\bar{s}$ pair, which implies that these structures are exotic candidates consisting of $cs\bar{c}\bar{s}$ constituent, such as multi-quark state, hadronic molecule, or hybrid.

Recently, the LHCb Collaboration reported evidence of a near-threshold peaking structure referred to as $X\left(3960\right)$ and another structure $X_{0}\left(4140\right)$ observed in the $D_{s}^{+}D_{s}^{-}$ invariant mass spectrum in the decay $B^{+}\to D_{s}^{+}D_{s}^{-}K^{+}$~\cite{LHCb:2022vsv}.
Before the observation of $X\left(3960\right)$, a plethora of exotic structures, that are good candidates of $cs\bar{c}\bar{s}$ tetraquarks, have been found, such as $X\left(4140\right)$~\cite{CDF:2009jgo,CMS:2013jru,D0:2013jvp,LHCb:2016axx}, $X\left(4274\right)$~\cite{CDF:2011pep,LHCb:2016axx}, $X\left(4350\right)$~\cite{Belle:2009rkh}, $X\left(4500\right)$~\cite{LHCb:2016axx}, $X\left(4630\right)$~\cite{LHCb:2021uow}, $X\left(4685\right)$~\cite{LHCb:2021uow}, and $X\left(4700\right)$~\cite{LHCb:2016axx}.

These signals are peculiarly interesting mainly because their probable constituent quark contents can be $cs\bar{c}\bar{s}$ in the hidden charm sector where the relatively heavier mass of the strange quark may provide quite different information concerning the dynamics describing the $D_s^+$ and $D_s^-$ interactions at the hadronic level or describing the constituent quark interactions at the quark level if we compare such a system with exotic candidates, such as $X(3872)$.

On the theory side, the mass spectrum of the $cs\bar{c}\bar{s}$ system has been studied by various approaches, such as the potential quark model~\cite{Braaten:2014qka,Vijande:2007fc,Vijande:2007rf,Fernandez-Carames:2009qqw,Yang:2021zhe,Tiwari:2022azj}, QCD sum rule~\cite{Chen:2017dpy,Xin:2022bzt}, lattice QCD (LQCD)~\cite{Chiu:2006us,Bali:2011rd,Prelovsek:2013cra,Sadl:2021bme}, effective field theory~\cite{Hidalgo-Duque:2012rqv,Ke:2013gia,Cleven:2013sq,Dong:2021juy,Ge:2021sdq,Chen:2022dad}, and the chromomagnetic interaction (CMI) model~\cite{Wu:2018xdi}.

About the nature of the charmonium-like states, there are many explanations in the literature. To know about the experimental and theoretical status of these charmonium-like states, one can refer to the recent reviews~\cite{Dong:2021bvy,Ali:2017jda,Esposito:2016noz,Olsen:2017bmm,Lebed:2016hpi,Liu:2019zoy,Chen:2016qju,Brambilla:2019esw,Chen:2022asf,Guo:2017jvc}.

Recently, the LHCb Collaboration reported the first observation of a double-charm tetraquark $T_{cc}^+(3875)$ ($cc\bar{u}\bar{d}$) in the $D^0D^0\pi^+$ invariant mass distribution~\cite{LHCb:2021vvq,LHCb:2021auc}. Its quantum numbers favor $IJ^{P}=01^+$. The double-charm tetraquark systems were studied in the literature before the observation of $T_{cc}^+(3875)$ and the theoretical predictions are quite different~\cite{Zouzou:1986qh,Lipkin:1986dw,Heller:1986bt,Carlson:1987hh,Silvestre-Brac:1993zem,Pepin:1996id,Brink:1998as,Janc:2004qn,Zhang:2007mu,Barnea:2006sd,Vijande:2007fc,Navarra:2007yw,Manohar:1992nd,Carames:2011zz,Ohkoda:2012hv}. With the observation of $T_{cc}^+(3875)$, a lot of interpretations and studies have been done based on different scenarios, such as compact states~\cite{Kim:2022mpa,Chen:2022ros}, hadronic molecules~\cite{Feijoo:2021ppq,Dong:2021bvy,Dai:2021wxi,Albaladejo:2021vln,Du:2021zzh,Ke:2021rxd,Abreu:2022sra,Xin:2021wcr,Peng:2023lfw}, triangle singularity (TS) mechanism~\cite{Braaten:2022elw}, etc. It was also investigated by LQCD calculations~\cite{Padmanath:2022cvl,Chen:2022vpo,Lyu:2023xro,Du:2023hlu}, and in its production and decays~\cite{Hu:2021gdg,Ling:2021bir,Meng:2021jnw,Yan:2021wdl}. In Refs.~\cite{Qiu:2023uno} combined analyses of the $X(3872)$, $Z_c(3900)$ and $T_{cc}^+(3875)$ are presented, which reveal peculiar dynamics arising from the $D^{(*)}\bar{D}^{(*)}$ and $D^{(*)}D^{(*)}$ threshold interactions.

In this work, we investigate the mass spectra, possible decay modes, and inner structures of tetraquark states $cs\bar{c}\bar{s}$ and $cc\bar{s}\bar{s}$ in the nonrelativistic potential quark model (NRPQM). Our attention is paid to cases where strong couplings to the nearby $S$-wave decay channels can be identified. Taking into account the experimental fact that some of those exotic candidates are located in the vicinity of an $S$-wave threshold and the possibility of multi-quark states in QCD cannot be eliminated, we are interested in any clues for connecting the quark model multi-quark configurations and the open channel dynamics with each other. Such information should be helpful for unifying the quark model prescription of the multi-quark spectra with the threshold dynamics based on hadron-hadron interactions. As the first step forward, we will focus on the $1S$-wave spectra of the $cs\bar{c}\bar{s}$ and $cc\bar{s}\bar{s}$ tetraquarks. We mention in advance that because of the configuration mixings the $1S$-wave spectra have become rich enough for raising a lot of interesting questions.

As follows, we first give an introduction of the framework of the NRPQM in Sec.~\ref{Fram}. Then, the mass spectra and strong decay properties of the $1S$-wave $cs\bar{c}\bar{s}$ and $cc\bar{s}\bar{s}$ states will be presented in Sec.~\ref{Discussion} with discussions. A summary will be given in Sec.~\ref{SUM}.

\section{Model and method} \label{Fram}

\subsection{Mass spectrum}

\subsubsection{Hamiltonian}

The mass spectra of the tetraquarks are calculated within the NRPQM, which has been widely adopted
to deal with the mesons and baryon spectra. In this model, the Hamiltonian is given by
\begin{eqnarray}
H & = & \left(\underset{i=1}{\overset{4}{\sum}}m_{i}+T_{i}\right)-T_{cm}+\underset{i<j}{\sum}V_{ij}\left(r_{ij}\right),\label{eq:H}
\end{eqnarray}
where $m_{i}$ and $T_{i}$ stand for the constituent quark mass and kinetic energy of the $i$-th quark, respectively; $T_{cm}$ stands for the center-of-mass (c.m.) kinetic energy of the tetraquark system; $\boldsymbol{r}_{ij}\equiv|\boldsymbol{r}_{i}-\boldsymbol{r}_{j}|$ is the distance between the $i$-th and $j$-th quarks; and $V_{ij}\left(r_{ij}\right)$ stands for the effective potential between them. In this work the $V_{ij}\left(r_{ij}\right)$ adopts a widely used form~\cite{Eichten:1978tg,Godfrey:1985xj,Barnes:2005pb,Godfrey:2015dia,Godfrey:2004ya,Lakhina:2006fy,Lu:2016bbk,Li:2010vx,Deng:2016stx,Deng:2016ktl,Liu:2021rtn,Liu:2019zuc,Liu:2020lpw}:
\begin{eqnarray}
V_{ij}\left(r_{ij}\right) & = & -\frac{3}{16}\left(\boldsymbol{\lambda}_{i}\cdot\boldsymbol{\lambda}_{j}\right)\left(b_{ij}r_{ij}-\frac{4}{3}\frac{\alpha_{ij}}{r_{ij}}+c_{ij}\right)\\
& + & -\frac{\alpha_{ij}}{4}\left(\boldsymbol{\lambda}_{i}\cdot\boldsymbol{\lambda}_{j}\right)\left\{ \frac{\pi}{2}\frac{\sigma_{ij}^{3}e^{-\sigma_{ij}^{2}r_{ij}^{2}}}{\pi^{3/2}}\frac{16}{3m_{i}m_{j}}\left(\boldsymbol{S}_{i}\cdot\boldsymbol{S}_{j}\right)\right\},\nonumber
\end{eqnarray}
where $\boldsymbol{\lambda}_{i,j}$ are the color operators acting on the $i,j$-th quarks,
$\boldsymbol{S}_{i,j}$ represent the spin operators of the $i,j$-th quarks.
The parameters $b_{ij}$ and $c_{ij}$ denote the confinement potential strength and the zero point energy, respectively. And $\alpha_{ij}$ denote the the strong coupling for the OGE potential. 
It should be mentioned that the tensor and spin-orbit potential do not contribute to the $1S$-wave tetraquarks considered here.

Parameters adopted in this work are collected in the Table~\ref{tab:parameters:qqqq}, which are extracted by fitting the mass spectra of the mesons~\cite{ParticleDataGroup:2020ssz} as also shown in Table~\ref{tab:parameter:meson}. The same quark model parameters are adopted as in Refs.~\cite{Deng:2016stx,Li:2020xzs,Li:2019qsg}.

\begin{table}
\centering \caption{Quark model parameters used in this work.}
\label{tab:parameters:qqqq}
\tabcolsep=0.45 cm
\begin{tabular}{cr@{\extracolsep{0pt}.}lr@{\extracolsep{0pt}.}lr@{\extracolsep{0pt}.}lcc}
\hline
\hline
$m_{u/d}$ (GeV) & 0&350 & \multicolumn{2}{c}{} & \multicolumn{2}{c}{} \tabularnewline
$m_{s}$ (GeV) & 0&600 & \multicolumn{2}{c}{} & \multicolumn{2}{c}{} \tabularnewline
$m_{c}$   (GeV)& 1&483 & \multicolumn{2}{c}{} & \multicolumn{2}{c}{} \tabularnewline
$m_{b}$   (GeV) & 4&852 & \multicolumn{2}{c}{} & \multicolumn{2}{c}{} \tabularnewline
\hline
 & \multicolumn{2}{c}{$ss$} & \multicolumn{2}{c}{$cc$} & \multicolumn{2}{c}{$cs$} \tabularnewline
\hline
$b_{qq}$ $ (\mathrm{GeV}^{2})$ & 0&1350 & 0&1425 & 0&1230 \tabularnewline
$c_{qq}$   (GeV)& -0&5190 & 0&0 & -0&1745 \tabularnewline
$\sigma_{qq}$ (GeV)& 0&6000 & 1&1384 & 0&8000 \tabularnewline
$\alpha_{qq}$ & 0&7700 & 0&5461 & 0&6800 \tabularnewline
\hline
\hline
\end{tabular}
\end{table}

\begin{table}
\centering \caption{The theoretical mass spectra of $s\bar{s}$, $c\bar{c}$, and $c\bar{s}$ compared with 
the observations (labeled with Obs.) from the PDG~\cite{ParticleDataGroup:2022pth}. The unit of mass is MeV. }
\label{tab:parameter:meson} 
\begin{tabular}{lclclclc}
\hline
\hline
 &  \multicolumn{2}{c}{\underline{~~~~~~~~$s\bar{s}$~~~~~~~~}} &\multicolumn{2}{c}{\underline{~~~~~~~~$c\bar{c}$~~~~~~~~}}& \multicolumn{2}{c}{\underline{~~~~~~~~$c\bar{s}$~~~~~~~~}}\tabularnewline
$n^{2S+1}L_J$   & Obs. & Mass & Obs. & Mass & Obs. & Mass \tabularnewline
\hline
$1^{1}S_{0}$   &         &  797 & $\eta_{c}$(2984) & 2984 & $D_{s}(1968)$ & 1969 \tabularnewline
$1^{3}S_{1}$   & $\phi\left(1020\right)$ & 1017 & $J/\psi$(3097) & 3097 & $D_{s}^{*}(2112)$ & 2112 \tabularnewline
$2^{1}S_{0}$   & $\eta\left(1475\right)$ & 1619 & $\eta_{c}\left(2S\right)$(3638) & 3635 & $\cdot\cdot\cdot$& 2655 \tabularnewline
$2^{3}S_{1}$   & $\phi\left(1680\right)$ & 1699 & $\psi\left(3686\right)$ & 3679 & $D_{s1}^{*}\left(2700\right)$ & 2721 \tabularnewline
$1^{3}P_{0}$   & $f_{0}\left(1370\right)$ & 1373 & $\chi_{c0}(3415)$ & 3417 & $\cdot\cdot\cdot$& 2419 \tabularnewline
$1^{1}P_{1}\left(1P_{1}\right)$   & $h_{1}\left(1415\right)$ & 1462 & $h_{c}(3525)$ & 3522 & $\cdot\cdot\cdot$& 2518 \tabularnewline
$1^{3}P_{1}\left(1P'_{1}\right)$   & $f_{1}\left(1420\right)$ & 1492 & $\chi_{c1}(3511)$ & 3516 & $D_{s1}\left(2536\right)$ & 2534 \tabularnewline
$1^{3}P_{2}$  & $f_{2}^{\prime}\left(1525\right)$ & 1513 & $\chi_{c2}(3556)$ & 3552 & $D_{s2}^{*}\left(2573\right)$ & 2550 \tabularnewline
$1^{1}D_{2}\left(1D_{2}\right)$   &$\cdot\cdot\cdot$ & 1825 & $\cdot\cdot\cdot$& 3806 & $\cdot\cdot\cdot$& 2812 \tabularnewline
$1^{3}D_{1}$   &$\cdot\cdot\cdot$ & 1808 & $\psi\left(3770\right)$ & 3787 & $D_{s1}^{*}\left(2860\right)$ & 2836 \tabularnewline
$1^{3}D_{2}\left(1D'_{2}\right)$   & $\cdot\cdot\cdot$& 1840 & $\psi_{2}\left(3823\right)$ & 3807 &$\cdot\cdot\cdot$ & 2865 \tabularnewline
$1^{3}D_{3}$   & $\cdot\cdot\cdot$& 1822 & $\psi_{3}\left(3842\right)$ & 3812 & $D_{s3}^{*}\left(2860\right)$ & 2822 \tabularnewline
\hline
\hline
\end{tabular}
\end{table}

\subsubsection{Tetraquark configurations}

The wave function for a tetraquark system can be constructed as a product of the flavor, color, spin, and spatial configurations.

In the color space, there are two color-singlet bases $|6_{12}\bar{6}_{34}\rangle_{c}$ and $|\bar{3}_{12}3_{34}\rangle_{c}$, their wave functions are given by
\begin{eqnarray}
\left|6_{12}\bar{6}_{34}\right\rangle & = & \frac{1}{2\sqrt{6}}\left[\left(rb+br\right)\left(\bar{r}\bar{b}+\bar{b}\bar{r}\right)+\left(bg+gb\right)\left(\bar{b}\bar{g}+\bar{g}\bar{b}\right)\right.\nonumber \\
& &+\left(gr+rg\right)\left(\bar{g}\bar{r}+\bar{r}\bar{g}\right)\nonumber \\
& & \left.+2\left(rr\right)\left(\bar{r}\bar{r}\right)+2\left(bb\right)\left(\bar{b}\bar{b}\right)+2\left(gg\right)\left(\bar{g}\bar{g}\right)\right],\\
\left|\bar{3}_{12}3_{34}\right\rangle & = & \frac{1}{2\sqrt{3}}\left[\left(rb-br\right)\left(\bar{r}\bar{b}-\bar{b}\bar{r}\right)+\left(bg-gb\right)\left(\bar{b}\bar{g}-\bar{g}\bar{b}\right)\right.\nonumber \\
& & \left.+\left(gr-gr\right)\left(\bar{g}\bar{r}-\bar{r}\bar{g}\right)\right] \ ,
\end{eqnarray}
where $r \ (\bar{r})$, $b \ (\bar{b})$ and $g \ (\bar{g})$ denote the color wave function of a single quark (antiquark).

In the spin space, there are six spin bases, which are denoted by $\chi_{SS_{z}}^{S_{12}S_{34}}$. Where $S_{12}$ stands for the spin quantum number for the diquark ($q_{1}q_{2}$) or antidiquark ($\bar{q}_{1}\bar{q}_{2}$), while $S_{34}$ stands for the spin quantum number for the antidiquark ($\bar{q}_{3}\bar{q}_{4}$) or diquark ($q_{3}q_{4}$). $S$ is the total spin quantum number of the tetraquark $qq\bar{q}\bar{q}$ system, while $S_{z}$ stands for the third component of the total spin $\boldsymbol{S}$. The spin wave functions $\chi_{SS_{z}}^{S_{12}S_{34}}$ with a determined $S_{z}$ can be explicitly expressed as follows:
\begin{eqnarray}
\chi_{00}^{00} & = & \frac{1}{2}\left(\uparrow\downarrow\uparrow\downarrow-\downarrow\uparrow\uparrow\downarrow-\uparrow\downarrow\downarrow\uparrow+\downarrow\uparrow\downarrow\uparrow\right),\label{spin000}\\
\chi_{00}^{11} & = & \frac{1}{\sqrt{12}}(2\uparrow\uparrow\downarrow\downarrow-\uparrow\downarrow\uparrow\downarrow-\downarrow\uparrow\uparrow\downarrow-\uparrow\downarrow\downarrow\uparrow\nonumber\\
&&-\downarrow\uparrow\downarrow\uparrow+2\downarrow\downarrow\uparrow\uparrow), \\
\chi_{11}^{01} & = & \frac{1}{\sqrt{2}}\left(\uparrow\downarrow\uparrow\uparrow-\downarrow\uparrow\uparrow\uparrow\right),\\
\chi_{11}^{10} & = & \frac{1}{\sqrt{2}}\left(\uparrow\uparrow\uparrow\downarrow-\uparrow\uparrow\downarrow\uparrow\right),\\
\chi_{11}^{11} & = & \frac{1}{2}\left(\uparrow\uparrow\uparrow\downarrow+\uparrow\uparrow\downarrow\uparrow-\uparrow\downarrow\uparrow\uparrow-\downarrow\uparrow\uparrow\uparrow\right),\\
\chi_{22}^{11} & = & \uparrow\uparrow\uparrow\uparrow.
\end{eqnarray}

In the spatial space, the relative Jacobi coordinates with
the single-partial coordinates $\boldsymbol{r_{i}}$ ($i=1,2,3,4$) are defined by
\begin{eqnarray}
\boldsymbol{\xi}_{1} & \equiv & \boldsymbol{r_{1}}-\boldsymbol{r_{2}},\\
\boldsymbol{\xi}_{2} & \equiv & \boldsymbol{r_{3}}-\boldsymbol{r_{4}},\\
\boldsymbol{\xi}_{3} & \equiv & \frac{m_{1}\boldsymbol{r_{1}}+m_{2}\boldsymbol{r_{2}}}{m_{1}+m_{2}}-\frac{m_{3}\boldsymbol{r_{3}}+m_{4}\boldsymbol{r_{4}}}{m_{3}+m_{4}},\\
\boldsymbol{R} & \equiv & \frac{m_{1}\boldsymbol{r_{1}}+m_{2}\boldsymbol{r_{2}}+m_{3}\boldsymbol{r_{3}}+m_{4}\boldsymbol{r_{4}}}{m_{1}+m_{2}+m_{3}+m_{4}}.
\end{eqnarray}
Note that $\boldsymbol{\xi}_{1}$ and $\boldsymbol{\xi}_{2}$ stand for the relative Jacobi coordinates between two quarks $q_{1}$ and $q_{2}$ (or antiquarks $\bar{q}_{1}$ and $\bar{q}_{2}$), and two antiquarks $\bar{q}_{3}$ and $\bar{q}_{4}$ (or quarks $q_{3}$ and $q_{4}$), respectively. While $\boldsymbol{\xi}_{3}$ stands for the relative Jacobi coordinate between diquark $qq$ and anti-diquark $\bar{q}\bar{q}$. Using the above Jacobi coordinates, it is easy to obtain basis functions that have well-defined symmetry under permutations of the pairs ($12$) and ($34$)~\cite{Vijande:2009kj}. 

Considering the Pauli principle and color confinement for the $cs\bar{c}\bar{s}$ system, we have 12 $1S$ configurations, while for the $cc\bar{s}\bar{s}$ we have 4 $1S$ configurations. The spin-parity quantum numbers, notations, and wave functions for these configurations are presented in Table~\ref{tab:configuration:QsQs:1S} and Table~\ref{tab:configuration:QQss:1S}.

\begin{table}
\centering
\caption{Configurations of $1S$ states for $T_{cs\bar{c}\bar{s}}$ system. The subscripts and superscripts of the configurations are 
the spin quantum numbers and representations of the color SU(3) group, respectively. $\psi$, $\chi$, and $\left|6\bar{6}\right\rangle_{c}$/ $\left|\bar{3}3\right\rangle_{c}$ stand for the wave functions in the spatial, spin, and color spaces, respectively.  }\centering
\label{tab:configuration:QsQs:1S}\centering
\begin{tabular}{ccccc}
\hline
\hline
$J^{PC}$ & Configuration & \multicolumn{3}{c}{$\underline{\hspace{0.5cm}\mathrm{Wave~function}\hspace{0.5cm}}$} \tabularnewline
\hline
$0^{++}$ & $\left|\left(cs\right)_{0}^{6}\left(\bar{c}\bar{s}\right)_{0}^{\bar{6}}\right\rangle _{0}$ & $\psi_{000}^{1S}$ & $\chi_{00}^{00}$ & $\left|6\bar{6}\right\rangle_{c}$ \tabularnewline
& $\left|\left(cs\right)_{0}^{\bar{3}}\left(\bar{c}\bar{s}\right)_{0}^{3}\right\rangle _{0}$ & $\psi_{000}^{1S}$ & $\chi_{00}^{00}$ & $\left|\bar{3}3\right\rangle_{c}$ \tabularnewline
& $\left|\left(cs\right)_{1}^{6}\left(\bar{c}\bar{s}\right)_{1}^{\bar{6}}\right\rangle _{0}$ & $\psi_{000}^{1S}$ & $\chi_{00}^{11}$ & $\left|6\bar{6}\right\rangle_{c}$ \tabularnewline
& $\left|\left(cs\right)_{1}^{\bar{3}}\left(\bar{c}\bar{s}\right)_{1}^{3}\right\rangle _{0}$ & $\psi_{000}^{1S}$ & $\chi_{00}^{11}$ & $\left|\bar{3}3\right\rangle_{c}$ \tabularnewline
$1^{+-}$ & $\left|\left(cs\right)_{1}^{6}\left(\bar{c}\bar{s}\right)_{1}^{\bar{6}}\right\rangle _{1}$ & $\psi_{000}^{1S}$ & $\chi_{1m}^{11}$ & $\left|6\bar{6}\right\rangle_{c}$ \tabularnewline
& $\left|\left(cs\right)_{1}^{\bar{3}}\left(\bar{c}\bar{s}\right)_{1}^{3}\right\rangle _{1}$ & $\psi_{000}^{1S}$ & $\chi_{1m}^{11}$ & $\left|\bar{3}3\right\rangle_{c}$ \tabularnewline
& $\frac{1}{\sqrt{2}}\left(\left|\left(cs\right)_{1}^{6}\left(\bar{c}\bar{s}\right)_{0}^{\bar{6}}\right\rangle _{1}-\left|\left(cs\right)_{0}^{6}\left(\bar{c}\bar{s}\right)_{1}^{\bar{6}}\right\rangle _{1}\right)$ & $\psi_{000}^{1S}$ & $\frac{1}{\sqrt{2}}\left(\chi_{1m}^{10}-\chi_{1m}^{01}\right)$ & $\left|6\bar{6}\right\rangle_{c}$ \tabularnewline
& $\frac{1}{\sqrt{2}}\left(\left|\left(cs\right)_{1}^{\bar{3}}\left(\bar{c}\bar{s}\right)_{0}^{3}\right\rangle _{1}-\left|\left(cs\right)_{0}^{\bar{3}}\left(\bar{c}\bar{s}\right)_{1}^{3}\right\rangle _{1}\right)$ & $\psi_{000}^{1S}$ & $\frac{1}{\sqrt{2}}\left(\chi_{1m}^{10}-\chi_{1m}^{01}\right)$ & $\left|\bar{3}3\right\rangle_{c}$ \tabularnewline
$1^{++}$ & $\frac{1}{\sqrt{2}}\left(\left|\left(cs\right)_{1}^{6}\left(\bar{c}\bar{s}\right)_{0}^{\bar{6}}\right\rangle _{1}+\left|\left(cs\right)_{0}^{6}\left(\bar{c}\bar{s}\right)_{1}^{\bar{6}}\right\rangle _{1}\right)$ & $\psi_{000}^{1S}$ & $\frac{1}{\sqrt{2}}\left(\chi_{1m}^{10}+\chi_{1m}^{01}\right)$ & $\left|6\bar{6}\right\rangle_{c}$ \tabularnewline
& $\frac{1}{\sqrt{2}}\left(\left|\left(cs\right)_{1}^{\bar{3}}\left(\bar{c}\bar{s}\right)_{0}^{3}\right\rangle _{1}+\left|\left(cs\right)_{0}^{\bar{3}}\left(\bar{c}\bar{s}\right)_{1}^{3}\right\rangle _{1}\right)$ & $\psi_{000}^{1S}$ & $\frac{1}{\sqrt{2}}\left(\chi_{1m}^{10}+\chi_{1m}^{01}\right)$ & $\left|\bar{3}3\right\rangle_{c}$ \tabularnewline
$2^{++}$ & $\left|\left(cs\right)_{1}^{6}\left(\bar{c}\bar{s}\right)_{1}^{\bar{6}}\right\rangle _{2}$ & $\psi_{000}^{1S}$ & $\chi_{2m}^{11}$ & $\left|6\bar{6}\right\rangle_{c}$ \tabularnewline
& $\left|\left(cs\right)_{1}^{\bar{3}}\left(\bar{c}\bar{s}\right)_{1}^{3}\right\rangle _{2}$ & $\psi_{000}^{1S}$ & $\chi_{2m}^{11}$ & $\left|\bar{3}3\right\rangle_{c}$ \tabularnewline
\hline
\hline
\end{tabular}
\end{table}

\begin{table}
\centering
\caption{Configurations of $1S$ states for $T_{cc\bar{s}\bar{s}}$ system. The caption is the same as that of Table~\ref{tab:configuration:QsQs:1S}.}\centering
\label{tab:configuration:QQss:1S}\centering
\tabcolsep=0.45 cm
\begin{tabular}{ccccc}
\hline
\hline
$J^{P}$ & Configuration & \multicolumn{3}{c}{$\underline{\hspace{0.2cm}\mathrm{Wave~function}\hspace{0.2cm}}$} \tabularnewline
\hline
$0^{+}$ & $\left|\left\{ cc\right\} _{0}^{6}\left\{ \bar{s}\bar{s}\right\} _{0}^{\bar{6}}\right\rangle _{0}$ & $\psi_{000}^{1S}$ & $\chi_{00}^{00}$ & $\left|6\bar{6}\right\rangle_{c}$ \tabularnewline
& $\left|\left\{ cc\right\} _{1}^{\bar{3}}\left\{ \bar{s}\bar{s}\right\} _{1}^{3}\right\rangle _{0}$ & $\psi_{000}^{1S}$ & $\chi_{00}^{11}$ & $\left|\bar{3}3\right\rangle_{c}$ \tabularnewline
$1^{+}$ & $\left|\left\{ cc\right\} _{1}^{\bar{3}}\left\{ \bar{s}\bar{s}\right\} _{1}^{3}\right\rangle _{1}$ & $\psi_{000}^{1S}$ & $\chi_{11}^{11}$ & $\left|\bar{3}3\right\rangle_{c}$ \tabularnewline
$2^{+}$ & $\left|\left\{ cc\right\} _{1}^{\bar{3}}\left\{ \bar{s}\bar{s}\right\} _{1}^{3}\right\rangle _{2}$ & $\psi_{000}^{1S}$ & $\chi_{22}^{11}$ & $\left|\bar{3}3\right\rangle_{c}$ \tabularnewline
\hline
\hline
\end{tabular}
\end{table}

\subsubsection{Numerical method}

With the wave functions for all the configurations, the mass matrix elements of the Hamiltonian can be worked out.
To solve the four-body problem accurately, we adopt the explicitly correlated Gaussian method~\cite{Mitroy:2013eom,Varga:1995dm}. It is a well-established variational method to solve quantum-mechanical few-body problems in molecular, atomic, and nuclear physics. The trail wave function of the tetraquark states without any spatial excitations in the coordinate space is expanded in terms of correlated Gaussian basis. Such a basis function can be written as
\begin{equation}
\psi=\exp\left[-\sum_{i<j=1}^{4}\frac{1}{2d_{ij}^{2}}(\boldsymbol{r}_{i}-\boldsymbol{r}_{j})^{2}\right]. \label{spatial function1-1-1}
\end{equation}
For a tetraquark system $cs\bar{c}\bar{s}$ with zero angular momentum, $d_{ij}$ are adjustable parameters. We can take $d_{12}=d_{34}\equiv a$, $d_{13}\equiv b$, $d_{24}\equiv c$ and $d_{14}=d_{23}\equiv e$. For a tetraquark system $cc\bar{s}\bar{s}$ with zero angular momentum, we can take $d_{12}=a$, $d_{34}\equiv b$ and $d_{13}=d_{24}=d_{14}=d_{23}\equiv c$. It is convenient to use a set of the Jacobi coordinates $\xi=\left(\xi_{1},\xi_{2},\xi_{3}\right)$, instead of the relative distance vectors $(\boldsymbol{r}_{i}-\boldsymbol{r}_{j})$. Then the correlated Gaussian basis function can be rewritten as
\begin{equation}
\psi\left(\xi,A\right)=\exp\left(-\sum_{i,j}A_{ij}\boldsymbol{\xi}_{i}\cdot\boldsymbol{\xi}_{j}\right)\equiv\exp\left(-\tilde{\boldsymbol{\xi}}A\boldsymbol{\xi}\right),\label{spatial function1-1}
\end{equation}
and $A$ is a $3\times3$ symmetric positive-definite matrix whose elements are variational parameters.

The coordinate part of the trial wave function $\psi\left(\xi,A\right)$ can be formed as a linear combination of the correlated Gaussians
\begin{eqnarray}
\psi\left(\xi,A\right)=\sum_{k=1}^{\mathcal{N}}c_{k}G\left(\xi,A_{k}\right).
\end{eqnarray}
The accuracy of the trial function depends on the length of the expansion $\mathcal{N}$ and the nonlinear parameters $A_{k}$. In our calculations, following the method of Ref.~\cite{Hiyama:2003cu}, we let the variational parameters form a geometric progression. For example, for a variational parameter $d$, we take
\begin{eqnarray}
d_{n}=d_{1}q^{n-1}(n=1,\cdot\cdot\cdot,n_{d}^{max}).
\end{eqnarray}
There are three parameters $\{d_{1},d_{n_{d}^{max}},n^{max}\}$ to be determined through the variation method. The length of the expansion $\mathcal{N}$ is determined to by $\mathcal{N}=n_{a}^{max}n_{b}^{max}n_{c}^{max}n_{e}^{max}$. In this work, stable solutions are obtained with $n_{a}^{max}=n_{b}^{max}=n_{c}^{max}=n_{e}^{max}\equiv n^{max}=5$.

\begin{figure}
\centering
\label{fig:stable:cscs}
\includegraphics[width=1\linewidth]{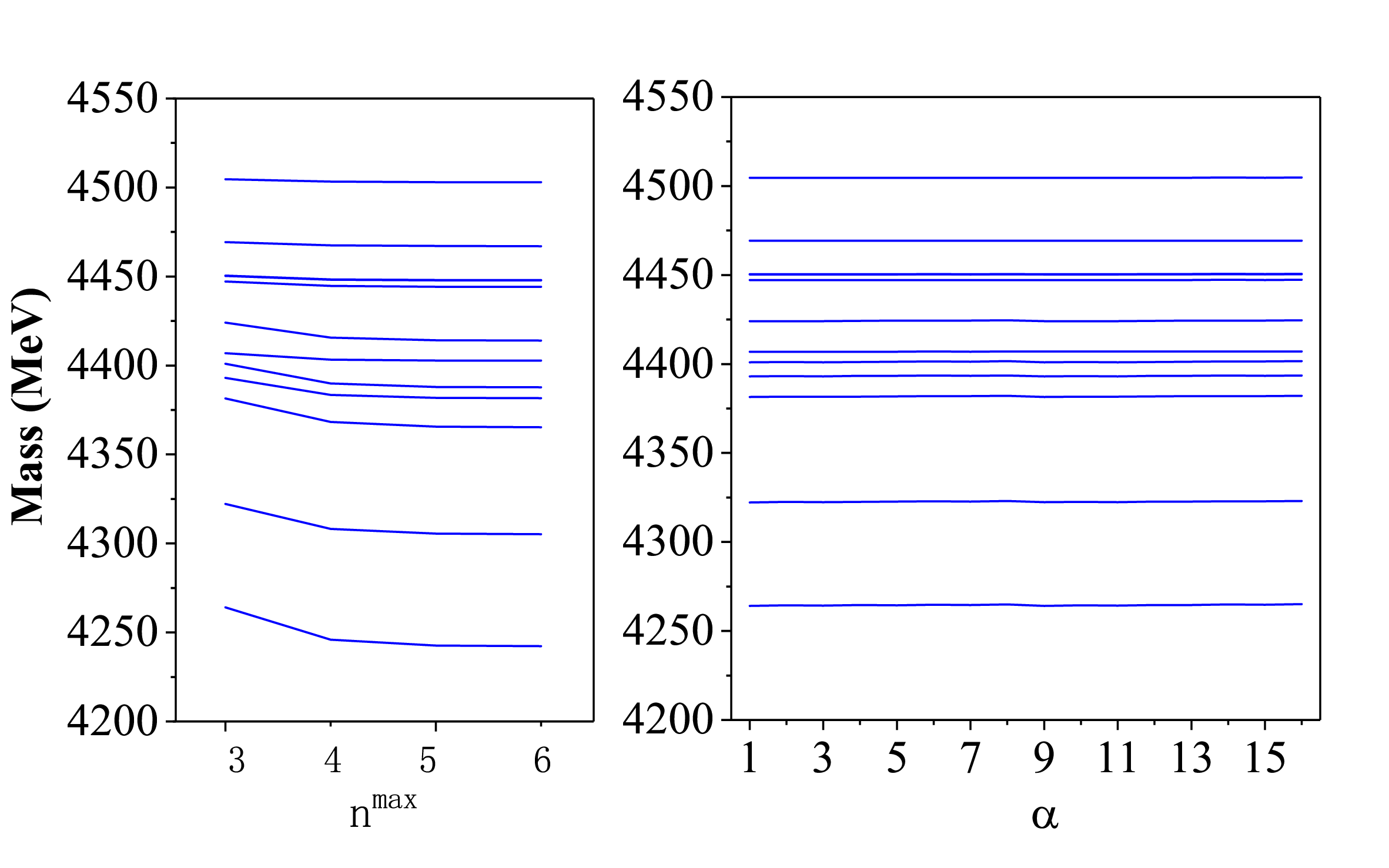}
\caption{Predicted masses of 12 $1S$-wave $T_{cs\bar{c}\bar{s}}$ configurations as a function of the scaling factor $n^{max}$ and $\alpha$.}
\end{figure}

The numerical results should be independent of the parameters $\{d_{1},d_{n_{d}^{max}},n_{d}^{max}\}$. 
To confirm this point, we scale the basis number parameter $n^{max}$ as $n^{max}=3\to 6$. As an example, we plot the masses of 12 $1S$-wave $T_{\left(cs\bar{c}\bar{s}\right)}$ configurations as a function of $n^{max}$ in Fig.~\ref{fig:stable:cscs}. 
It is found that the numerical results are nearly independent of $n^{max}$ when it is large enough. 
On the other hand, to see the $d_{n}$ independence of the numerical results, as commonly dealt with in the literature~\cite{Hiyama:2005cf,Hiyama:2018ukv,Meng:2019fan} we scale the parameter $d_{1}$ of the basis functions as $d_{1}\to\alpha d_{1}$. The mass of a $T_{cs\bar{c}\bar{s}}$ state should be stable at a resonance energy insensitive to the scaling parameter $\alpha$. As an example, we plot the masses of 12 $1S$-wave $T_{\left(cs\bar{c}\bar{s}\right)}$ configurations as a function of the scaling factor $\alpha$ in Fig.~\ref{fig:stable:cscs}. It is found that the numerical results are nearly independent of the scaling factor $\alpha$. The stabilization of other states predicted in this work has also been examined by the same method.

With the mass matrix elements ready for every configuration, the mass of the tetraquark configuration and its spacial wave function can be determined by solving a generalized eigenvalue problem. The details can be found in our previous works~\cite{Liu:2019vtx,Liu:2019zuc}. The physical states can be obtained by diagonalizing the mass matrix of different configurations with the same $J^{PC}$ numbers.

\subsection{Rearrangement decay}

With the eigenstates obtained in the previous section, we can calculate the rearrangement decays of the $cs\bar{c}\bar{s}$ and $cc\bar{s}\bar{s}$ systems in a quark-exchange model~\cite{Barnes:2000hu}. The transition operators can be extracted from the quark-quark and quark-antiquark interactions $V_{ij}$ via the quark rearrangement. The decay amplitude $\mathcal{M}(A\to BC)$ of a tetraquark state is described by
\begin{eqnarray}
\mathcal{M}(A\to BC)=-\sqrt{(2\pi)^3}\sqrt{8M_AE_BE_C}\left\langle BC |\sum_{i<j} V_{ij}| A \right\rangle,
\end{eqnarray}
where $A$ stands for the initial tetraquark state, $BC$ stands for the final hadron pair. The potentials $V_{ij}$ between quarks are taken from the Hamiltonian. $M_A$ is the mass of the initial state, while $E_B$ and $E_C$ are the energies of the final states $B$ and $C$, respectively, in the initial-hadron-rest system.

This phenomenological model has been applied to the study of the hidden-charm decay properties for the multiquark states in the literature~\cite{Wang:2019spc,Xiao:2019spy,Wang:2020prk,Han:2022fup}. For simplicity, the wave functions of the initial and final state hadrons, i.e. $A,\ B, \ C$, are adopted in the form of single harmonic oscillator. They are determined by fitting the wave functions calculated from our potential model. The partial decay width of $A\to BC$ is given by
\begin{eqnarray}
\Gamma=\frac{1}{2J_A+1}\frac{|\boldsymbol{q}|}{8\pi M_A^2}\left|\mathcal{M}(A\to BC)\right|^2,
\end{eqnarray}
where $\boldsymbol{q}$ is the three-vector momentum of the final state $B$ or $C$ in the initial-hadron-rest
frame. We note that the rearrangement decays of the tetraquarks actually provide a dynamical description of the couplings of the tetraquarks to two color-singlet hadrons in the final state. It can play a role as the bare coupling of the elementary component of an exotic state to the constituent hadrons in a molecular picture if the coupling is via a near-threshold $S$-wave interaction.

\begin{table}
\centering
\caption{The average contributions of each part of the Hamiltonian to the $T_{cs\bar{c}\bar{s}\ }\left(1S\right)$ configurations. In the table, we define that $\frac{1}{\sqrt{2}}\left(\left|\left(cs\right)_{1}^{6}\left(\bar{c}\bar{s}\right)_{0}^{\bar{6}}\right\rangle _{1}\right)^{\pm}\equiv\frac{1}{\sqrt{2}}\left(\left|\left(cs\right)_{1}^{6}\left(\bar{c}\bar{s}\right)_{0}^{\bar{6}}\right\rangle _{1}\pm\left|\left(cs\right)_{0}^{6}\left(\bar{c}\bar{s}\right)_{1}^{\bar{6}}\right\rangle _{1}\right)$ and $\frac{1}{\sqrt{2}}\left(\left|\left(cs\right)_{1}^{\bar{3}}\left(\bar{c}\bar{s}\right)_{0}^{3}\right\rangle _{1}\right)^{\pm}\equiv\frac{1}{\sqrt{2}}\left(\left|\left(cs\right)_{1}^{\bar{3}}\left(\bar{c}\bar{s}\right)_{0}^{3}\right\rangle _{1}\pm\left|\left(cs\right)_{0}^{\bar{3}}\left(\bar{c}\bar{s}\right)_{1}^{3}\right\rangle _{1}\right)$.}
\label{tab:contributions:cscs:1S1}\centering
\begin{tabular}{ccccccr@{\extracolsep{0pt}.}lc}
\hline
\hline
$J^{PC}$ & State & Mass & $\left\langle T\right\rangle $ & $\left\langle V^{Lin}\right\rangle $ & $\left\langle V^{Coul}\right\rangle $ & \multicolumn{2}{c}{$\left\langle V^{SS}\right\rangle $} & $\left\langle V^{c}\right\rangle $ \tabularnewline
\hline
$0^{++}$ & $\left|\left(cs\right)_{0}^{6}\left(\bar{c}\bar{s}\right)_{0}^{\bar{6}}\right\rangle _{0}$ & 4388 & 773 & 740 & -854 & 18&13 & -455 \tabularnewline
& $\left|\left(cs\right)_{0}^{\bar{3}}\left(\bar{c}\bar{s}\right)_{0}^{3}\right\rangle _{0}$ & 4403 & 855 & 701 & -859 & -68&53 & -391 \tabularnewline
& $\left|\left(cs\right)_{1}^{6}\left(\bar{c}\bar{s}\right)_{1}^{\bar{6}}\right\rangle _{0}$ & 4243 & 954 & 670 & -951 & -140&51 & -455 \tabularnewline
& $\left|\left(cs\right)_{1}^{\bar{3}}\left(\bar{c}\bar{s}\right)_{1}^{3}\right\rangle _{0}$ & 4448 & 797 & 724 & -828 & -19&73 & -391 \tabularnewline
$1^{+-}$ & $\left|\left(cs\right)_{1}^{6}\left(\bar{c}\bar{s}\right)_{1}^{\bar{6}}\right\rangle _{1}$ & 4305 & 874 & 698 & -910 & -67&32 & -455 \tabularnewline
& $\left|\left(cs\right)_{1}^{\bar{3}}\left(\bar{c}\bar{s}\right)_{1}^{3}\right\rangle _{1}$ & 4467 & 774 & 734 & -815 & 0&34 & -391 \tabularnewline
& $\frac{1}{\sqrt{2}}\left(\left|\left(cs\right)_{1}^{6}\left(\bar{c}\bar{s}\right)_{0}^{\bar{6}}\right\rangle _{1}\right)^{-}$ & 4366 & 798 & 730 & -867 & -5&46 & -455 \tabularnewline
& $\frac{1}{\sqrt{2}}\left(\left|\left(cs\right)_{1}^{\bar{3}}\left(\bar{c}\bar{s}\right)_{0}^{3}\right\rangle _{1}\right)^{-}$ & 4444 & 801 & 722 & -831 & -22&69 & -391 \tabularnewline
$1^{++}$ & $\frac{1}{\sqrt{2}}\left(\left|\left(cs\right)_{1}^{6}\left(\bar{c}\bar{s}\right)_{0}^{\bar{6}}\right\rangle _{1}\right)^{+}$ & 4382 & 785 & 735 & -859 & 10&93 & -455 \tabularnewline
& $\frac{1}{\sqrt{2}}\left(\left|\left(cs\right)_{1}^{\bar{3}}\left(\bar{c}\bar{s}\right)_{0}^{3}\right\rangle _{1}\right)^{+}$ & 4448 & 798 & 723 & -829 & -18&96 & -392 \tabularnewline
$2^{++}$ & $\left|\left(cs\right)_{1}^{6}\left(\bar{c}\bar{s}\right)_{1}^{\bar{6}}\right\rangle _{2}$ & 4414 & 746 & 751 & -837 & 43&23 & -455 \tabularnewline
& $\left|\left(cs\right)_{1}^{\bar{3}}\left(\bar{c}\bar{s}\right)_{1}^{3}\right\rangle _{2}$ & 4503 & 732 & 753 & -792 & 35&39 & -391 \tabularnewline
\hline
\hline
\end{tabular}
\end{table}

\begin{table}
\centering
\caption{The average contributions of each part of the Hamiltonian to the $T_{cc\bar{s}\bar{s}\ }\left(1S\right)$ configurations.}
\label{tab:contributions:ccss:1S1}
\begin{tabular}{ccccccr@{\extracolsep{0pt}.}lc}
\hline
\hline
$J^{P}$ & $\ensuremath{\hspace{0.9 cm}}$ State $\ensuremath{\hspace{0.9 cm}}$ & Mass & $\left\langle T\right\rangle $ & $\left\langle V^{Lin}\right\rangle $ & $\left\langle V^{Coul}\right\rangle $ & \multicolumn{2}{c}{$\left\langle V^{SS}\right\rangle $} & $\left\langle V^{c}\right\rangle $ \tabularnewline
\hline
$0^{+}$ & $\left|\left\{ cc\right\} _{0}^{6}\left\{ \bar{s}\bar{s}\right\} _{0}^{\bar{6}}\right\rangle _{0}$ & 4465 & 735 & 682 & -831 & $\ $19&99 & -306 \tabularnewline
& $\left|\left\{ cc\right\} _{1}^{\bar{3}}\left\{ \bar{s}\bar{s}\right\} _{1}^{3}\right\rangle _{0}$ & 4432 & 802 & 740 & -831 & -10&54 & -434 \tabularnewline
$1^{+}$ & $\left|\left\{ cc\right\} _{1}^{\bar{3}}\left\{ \bar{s}\bar{s}\right\} _{1}^{3}\right\rangle _{1}$ & 4449 & 780 & 750 & -820 & 7&11 & -434 \tabularnewline
$2^{+}$ & $\left|\left\{ cc\right\} _{1}^{\bar{3}}\left\{ \bar{s}\bar{s}\right\} _{1}^{3}\right\rangle _{2}$ & 4481 & 740 & 769 & -799 & 38&16 & -434 \tabularnewline
\hline
\hline
\end{tabular}
\end{table}

\begin{figure}
\centering
\label{fig:mass:cscs}
\includegraphics[width=1\linewidth]{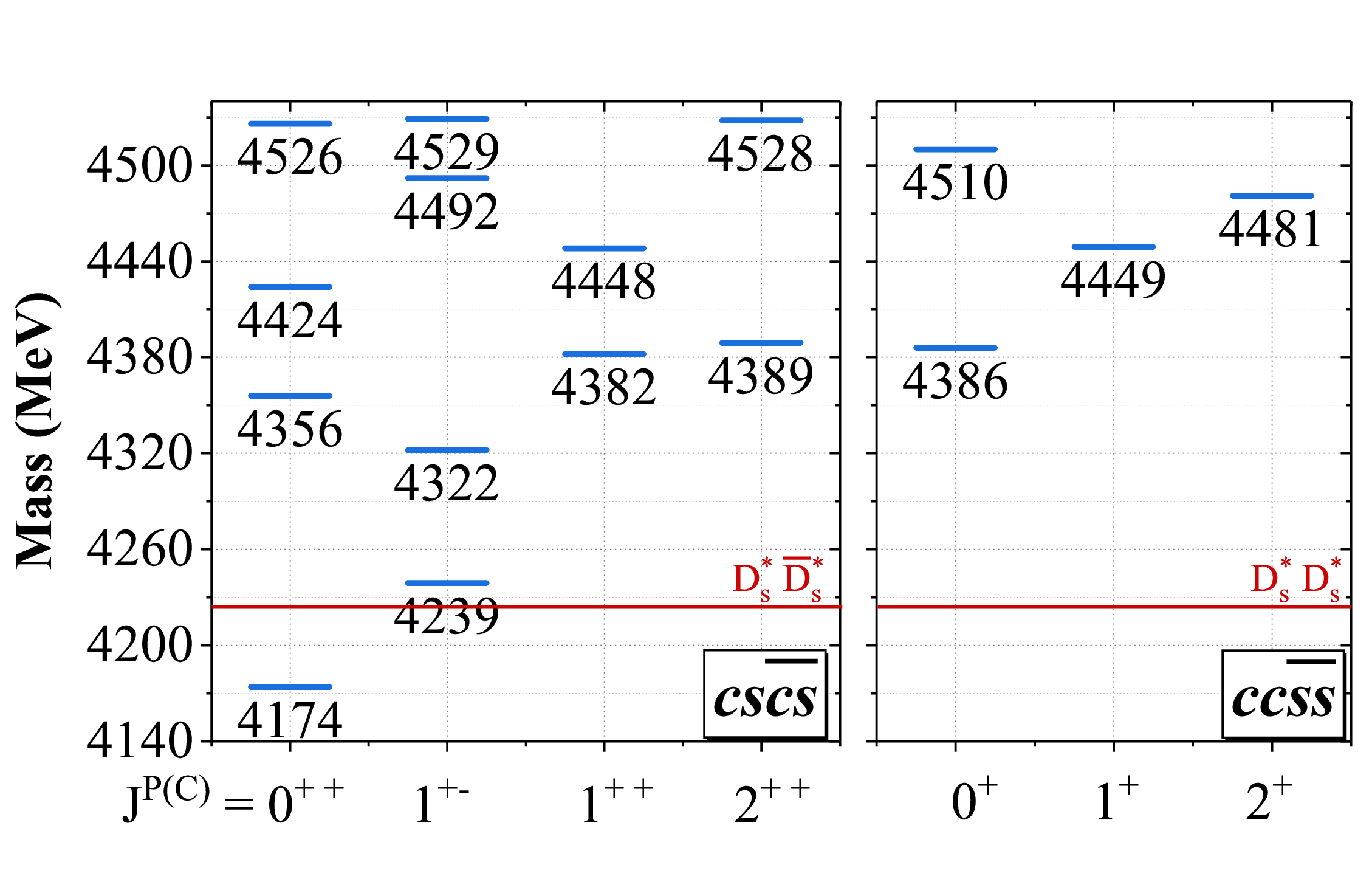}
\caption{Predicted mass spectra of $1S$ states for the $cs\bar{c}\bar{s}$ and $T_{cc\bar{s}\bar{s}}$ systems.}
\end{figure}

\begin{table*}
\centering
\caption{Predicted mass spectra of $1S$ states for the $T_{cs\bar{c}\bar{s}}$ system.}
\label{tab:mass:cscs:1S}
\tabcolsep=0.35 cm
\begin{tabular}{ccccc}
\hline
\hline
$J^{PC}$ & State & $\left\langle H\right\rangle $~(MeV) & Mass~(MeV) & Eigenvector \tabularnewline
\hline
$0^{++}$ & $\begin{array}{c}
\left|\left(cs\right)_{0}^{6}\left(\bar{c}\bar{s}\right)_{0}^{\bar{6}}\right\rangle _{0}\\
\left|\left(cs\right)_{0}^{\bar{3}}\left(\bar{c}\bar{s}\right)_{0}^{3}\right\rangle _{0}\\
\left|\left(cs\right)_{1}^{6}\left(\bar{c}\bar{s}\right)_{1}^{\bar{6}}\right\rangle _{0}\\
\left|\left(cs\right)_{1}^{\bar{3}}\left(\bar{c}\bar{s}\right)_{1}^{3}\right\rangle _{0}
\end{array}$ & $\left(\begin{array}{rrrr}
4388 & -58 & -15 & -71\\
-58 & 4403 & -81 & -3\\
-15 & -81 & 4243 & -76\\
-71 & -3 & -76 & 4448
\end{array}\right)$ & $\left(\begin{array}{r}
4174\\
4356\\
4424\\
4526
\end{array}\right)$ & $\left(\begin{array}{rrrr}
0.26 & 0.37 & 0.84 & 0.31\\
0.82 & 0.24 & -0.46 & 0.26\\
0.11 & -0.80 & 0.11 & 0.58\\
0.50 & -0.40 & 0.28 & -0.71
\end{array}\right)$ \tabularnewline
$1^{+-}$ & $\begin{array}{c}
\left|\left(cs\right)_{1}^{6}\left(\bar{c}\bar{s}\right)_{1}^{\bar{6}}\right\rangle _{1}\\
\left|\left(cs\right)_{1}^{\bar{3}}\left(\bar{c}\bar{s}\right)_{1}^{3}\right\rangle _{1}\\
\frac{1}{\sqrt{2}}\left(\left|\left(cs\right)_{1}^{6}\left(\bar{c}\bar{s}\right)_{0}^{\bar{6}}\right\rangle _{1}-\left|\left(cs\right)_{0}^{6}\left(\bar{c}\bar{s}\right)_{1}^{\bar{6}}\right\rangle _{1}\right)\\
\frac{1}{\sqrt{2}}\left(\left|\left(cs\right)_{1}^{\bar{3}}\left(\bar{c}\bar{s}\right)_{0}^{3}\right\rangle _{1}-\left|\left(cs\right)_{0}^{\bar{3}}\left(\bar{c}\bar{s}\right)_{1}^{3}\right\rangle _{1}\right)
\end{array}$ & $\left(\begin{array}{rrrr}
4305 & -68 & -28 & -23\\
-68 & 4467 & -21 & -10\\
-28 & -21 & 4366 & -117\\
-23 & -10 & -117 & 4444
\end{array}\right)$ & $\left(\begin{array}{r}
4239\\
4322\\
4492\\
4529
\end{array}\right)$ & $\left(\begin{array}{rrrr}
0.66 & 0.27 & 0.57 & 0.41\\
0.68 & 0.20 & -0.57 & -0.41\\
0.33 & -0.94 & -0.03 & 0.12\\
0.04 & -0.11 & 0.58 & -0.80
\end{array}\right)$ \tabularnewline
$1^{++}$ & $\begin{array}{c}
\frac{1}{\sqrt{2}}\left(\left|\left(cs\right)_{1}^{6}\left(\bar{c}\bar{s}\right)_{0}^{\bar{6}}\right\rangle _{1}+\left|\left(cs\right)_{0}^{6}\left(\bar{c}\bar{s}\right)_{1}^{\bar{6}}\right\rangle _{1}\right)\\
\frac{1}{\sqrt{2}}\left(\left|\left(cs\right)_{1}^{\bar{3}}\left(\bar{c}\bar{s}\right)_{0}^{3}\right\rangle _{1}+\left|\left(cs\right)_{0}^{\bar{3}}\left(\bar{c}\bar{s}\right)_{1}^{3}\right\rangle _{1}\right)
\end{array}$ & $\left(\begin{array}{rr}
4382 & -2\\
-2 & 4448
\end{array}\right)$ & $\left(\begin{array}{r}
4382\\
4448
\end{array}\right)$ & $\left(\begin{array}{rr}
1.00 & 0.03\\
0.03 & -1.00
\end{array}\right)$ \tabularnewline
$2^{++}$ & $\begin{array}{c}
\left|\left(cs\right)_{1}^{6}\left(\bar{c}\bar{s}\right)_{1}^{\bar{6}}\right\rangle _{2}\\
\left|\left(cs\right)_{1}^{\bar{3}}\left(\bar{c}\bar{s}\right)_{1}^{3}\right\rangle _{2}
\end{array}$ & $\left(\begin{array}{rr}
4414 & -54\\
-54 & 4503
\end{array}\right)$ & $\left(\begin{array}{r}
4389\\
4528
\end{array}\right)$ & $\left(\begin{array}{rr}
0.90 & 0.43\\
0.43 & -0.90
\end{array}\right)$ \tabularnewline
\hline
\hline
\end{tabular}
\end{table*}

\begin{table}
\centering
\caption{Predicted mass spectra of $1S$ states for the $T_{cc\bar{s}\bar{s}}$ system.}
\label{tab:mass:ccss:1S}
\begin{tabular}{ccccc}
\hline
\hline
$J^{P}$ & State & $\left\langle H\right\rangle $~(MeV) & Mass~(MeV) & Eigenvector \tabularnewline
\hline
$0^{+}$ & $\begin{array}{c}
\left|\left\{ cc\right\} _{0}^{6}\left\{ \bar{s}\bar{s}\right\} _{0}^{\bar{6}}\right\rangle _{0}\\
\left|\left\{ cc\right\} _{1}^{\bar{3}}\left\{ \bar{s}\bar{s}\right\} _{1}^{3}\right\rangle _{0}
\end{array}$ & $\left(\begin{array}{rr}
4465 & -60\\
-60 & 4432
\end{array}\right)$ & $\left(\begin{array}{r}
4386\\
4510
\end{array}\right)$ & $\left(\begin{array}{rr}
-0.61 & -0.80\\
-0.80 & 0.61
\end{array}\right)$ \tabularnewline
$1^{+}$ & $\left|\left\{ cc\right\} _{1}^{\bar{3}}\left\{ \bar{s}\bar{s}\right\} _{1}^{3}\right\rangle _{1}$ & 4449 & 4449 & 1 \tabularnewline
$2^{+}$ & $\left|\left\{ cc\right\} _{1}^{\bar{3}}\left\{ \bar{s}\bar{s}\right\} _{1}^{3}\right\rangle _{2}$ & 4481 & 4481 & 1 \tabularnewline
\hline
\hline
\end{tabular}
\end{table}

\begin{table*}
\centering
\caption{The proportions of different color configurations and the root mean square radius for each state of $cs\bar{c}\bar{s}$. And $\langle\boldsymbol{r}_{12}^{2}\rangle^{\frac{1}{2}}$=$\langle\boldsymbol{r}_{34}^{2}\rangle^{\frac{1}{2}}$, $\langle\boldsymbol{r}_{14}^{2}\rangle^{\frac{1}{2}}$=$\langle\boldsymbol{r}_{23}^{2}\rangle^{\frac{1}{2}}$.}\centering
\label{tab:squareradius:ABAB}
\tabcolsep=0.281 cm
\begin{tabular}{crrrrccccccc}
\hline \hline
State & $\left|6\bar{6}\right\rangle $ & $\left|\bar{3}3\right\rangle $ & $\left|11\right\rangle $ & $\left|88\right\rangle $ & $\langle\boldsymbol{r}_{12}^{2}\rangle^{\frac{1}{2}}$ & $\langle\boldsymbol{r}_{13}^{2}\rangle^{\frac{1}{2}}$ & $\langle\boldsymbol{r}_{24}^{2}\rangle^{\frac{1}{2}}$ & $\langle\boldsymbol{r}_{14}^{2}\rangle^{\frac{1}{2}}$ & $\langle\boldsymbol{r}_{12-34}^{2}\rangle^{\frac{1}{2}}$ & $\langle\boldsymbol{r}_{13-24}^{2}\rangle^{\frac{1}{2}}$ & $\langle\boldsymbol{r}_{14-23}^{2}\rangle^{\frac{1}{2}}$ \tabularnewline
\hline
$T_{cs\bar{c}\bar{s}\ 0^{++}}\left(4174\right)$ & 77$\%$ & 23$\%$ & 92$\%$ & 8$\%$ & 0.64 & 0.50 & 0.66 & 0.60 & 0.38 & 0.46 & 0.42 \tabularnewline
$T_{cs\bar{c}\bar{s}\ 0^{++}}\left(4356\right)$ & 88$\%$ & 12$\%$ & 70$\%$ & 30$\%$ & 0.69 & 0.52 & 0.70 & 0.64 & 0.38 & 0.51 & 0.45 \tabularnewline
$T_{cs\bar{c}\bar{s}\ 0^{++}}\left(4424\right)$ & 2$\%$ & 98$\%$ & 32$\%$ & 68$\%$ & 0.59 & 0.54 & 0.70 & 0.63 & 0.45 & 0.42 & 0.41 \tabularnewline
$T_{cs\bar{c}\bar{s}\ 0^{++}}\left(4526\right)$ & 34$\%$ & 66$\%$ & 7$\%$ & 93$\%$ & 0.63 & 0.53 & 0.69 & 0.63 & 0.42 & 0.45 & 0.43 \tabularnewline
$T_{cs\bar{c}\bar{s}\ 1^{+-}}\left(4239\right)$ & 76$\%$ & 24$\%$ & 98$\%$ & 2$\%$ & 0.67 & 0.51 & 0.68 & 0.63 & 0.39 & 0.49 & 0.44 \tabularnewline
$T_{cs\bar{c}\bar{s}\ 1^{+-}}\left(4322\right)$ & 79$\%$ & 21$\%$ & 95$\%$ & 5$\%$ & 0.67 & 0.51 & 0.68 & 0.63 & 0.38 & 0.49 & 0.43 \tabularnewline
$T_{cs\bar{c}\bar{s}\ 1^{+-}}\left(4492\right)$ & 10$\%$ & 90$\%$ & 8$\%$ & 92$\%$ & 0.63 & 0.55 & 0.72 & 0.65 & 0.45 & 0.45 & 0.43 \tabularnewline
$T_{cs\bar{c}\bar{s}\ 1^{+-}}\left(4529\right)$ & 35$\%$ & 65$\%$ & 1$\%$ & 99$\%$ & 0.63 & 0.53 & 0.70 & 0.64 & 0.43 & 0.46 & 0.42 \tabularnewline
$T_{cs\bar{c}\bar{s}\ 1^{++}}\left(4382\right)$ & 100$\%$ & & 70$\%$ & 30$\%$ & 0.71 & 0.53 & 0.72 & 0.63 & 0.38 & 0.50 & 0.47 \tabularnewline
$T_{cs\bar{c}\bar{s}\ 1^{++}}\left(4448\right)$ & & 100$\%$ & 30$\%$ & 70$\%$ & 0.60 & 0.55 & 0.72 & 0.64 & 0.46 & 0.43 & 0.42 \tabularnewline
$T_{cs\bar{c}\bar{s}\ 2^{++}}\left(4389\right)$ & 82$\%$ & 18$\%$ & 98$\%$ & 2$\%$ & 0.71 & 0.54 & 0.73 & 0.67 & 0.40 & 0.52 & 0.47 \tabularnewline
$T_{cs\bar{c}\bar{s}\ 2^{++}}\left(4528\right)$ & 18$\%$ & 82$\%$ & 2$\%$ & 98$\%$ & 0.65 & 0.57 & 0.75 & 0.67 & 0.47 & 0.47 & 0.45 \tabularnewline
\hline \hline
\end{tabular}
\centering \caption{The proportions of different color configurations and the root mean
square radius for each state of $cc\bar{s}\bar{s}$. And $\langle\boldsymbol{r}_{13}^{2}\rangle^{\frac{1}{2}}$=$\langle\boldsymbol{r}_{24}^{2}\rangle^{\frac{1}{2}}$=$\langle\boldsymbol{r}_{14}^{2}\rangle^{\frac{1}{2}}$=$\langle\boldsymbol{r}_{23}^{2}\rangle^{\frac{1}{2}}$,
$\langle\boldsymbol{r}_{13-24}^{2}\rangle^{\frac{1}{2}}$=$\langle\boldsymbol{r}_{14-23}^{2}\rangle^{\frac{1}{2}}$.}
\label{tab:squareradius:AABB} \tabcolsep=0.45 cm
\begin{tabular}{crrrrccccc}
\hline \hline
State & $\left|6\bar{6}\right\rangle $ & $\left|\bar{3}3\right\rangle $ & $\left|11\right\rangle $ & $\left|88\right\rangle $ & $\langle\boldsymbol{r}_{12}^{2}\rangle^{\frac{1}{2}}$ & $\langle\boldsymbol{r}_{34}^{2}\rangle^{\frac{1}{2}}$ & $\langle\boldsymbol{r}_{13}^{2}\rangle^{\frac{1}{2}}$ & $\langle\boldsymbol{r}_{12-34}^{2}\rangle^{\frac{1}{2}}$ & $\langle\boldsymbol{r}_{13-24}^{2}\rangle^{\frac{1}{2}}$ \tabularnewline
\hline
$T_{cc\bar{s}\bar{s}\ 0^{+}}\left(4386\right)$ & 36$\%$ & 64$\%$ & 45$\%$ & 55$\%$ & 0.57 & 0.74 & 0.65 & 0.45 & 0.45 \tabularnewline
$T_{cc\bar{s}\bar{s}\ 0^{+}}\left(4510\right)$ & 64$\%$ & 36$\%$ & 55$\%$ & 45$\%$ & 0.59 & 0.76 & 0.65 & 0.42 & 0.48 \tabularnewline
$T_{cc\bar{s}\bar{s}\ 1^{+}}\left(4449\right)$ & & 100$\%$ & 33$\%$ & 67$\%$ & 0.51 & 0.69 & 0.65 & 0.49 & 0.41 \tabularnewline
$T_{cc\bar{s}\bar{s}\ 2^{+}}\left(4481\right)$ & & 100$\%$ & 33$\%$ & 67$\%$ & 0.51 & 0.71 & 0.68 & 0.51 & 0.42 \tabularnewline
\hline \hline
\end{tabular}
\end{table*}

\section{Results and discussions} \label{Discussion}

\subsection{Mass spectra and quark configurations}

The predicted masses of each configuration for the $cs\bar{c}\bar{s}$ and $cc\bar{s}\bar{s}$ systems have been listed in 
Table~\ref{tab:contributions:cscs:1S1} and~\ref{tab:contributions:ccss:1S1}, respectively. The contributions from each part 
of the Hamiltonian to these configurations are further analyzed. The results are listed in Table~\ref{tab:contributions:cscs:1S1} 
and~\ref{tab:contributions:ccss:1S1} as well. It shows that the averaged kinetic energy $\langle T\rangle$, the linear confining 
potential $\langle V^{Lin}\rangle$, and the Coulomb potential $\langle V^{Coul}\rangle$ have the same order of magnitude. 
Furthermore, it is found that the spin-spin interaction plays an important role in the configuration 
$\left|\left(cs\right)_{1}^{6}\left(\bar{c}\bar{s}\right)_{1}^{\bar{6}}\right\rangle _{0}$. The predicted mass for this 
configuration is notably ($\sim60-260$ MeV) smaller than the other configurations due to the strong attractive spin-spin 
interactions $\langle V^{SS}\rangle\simeq-140$ MeV. 

After considering configuration mixing, one can obtain the physical states. The predicted mass spectra for the $cs\bar{c}\bar{s}$ and $cc\bar{s}\bar{s}$ systems have been given in Tables~\ref{tab:mass:cscs:1S} and~\ref{tab:mass:ccss:1S} and also shown in Fig.~\ref{fig:mass:cscs}. The masses of the $1S$-wave $cs\bar{c}\bar{s}$ and $cc\bar{s}\bar{s}$ states are predicted to be in the range of $\sim4.1-4.6$ GeV and $\sim4.3-4.6$ GeV, respectively. One can see that the physical states are usually mixtures of two different color configurations $|6\bar{6}\rangle_c$ and
$|\bar{3}3\rangle_c$. For the physical states with $J^P=0^{++}$, $1^{+-}$ and $2^{++}$, the mixing between different color configurations is particularly evident. The configuration mixing effects can cause notable mass shifts to the physical states.
It shows that our results of the $1S$-wave $cs\bar{c}\bar{s}$ states are much higher than other theoretical works~\cite{Li:2023wxm,Guo:2022crh,Lu:2016cwr}, but consistent with Ref.~\cite{Wu:2016gas}. Meanwhile, our results of the $1S$-wave $cc\bar{s}\bar{s}$ states are consistent with other theoretical works~\cite{Lu:2020rog,Ebert:2007rn,Zhang:2007mu}.

\subsection{Proportions of different color configurations}

It should be mentioned that except for the color configurations $|6\bar{6}\rangle_{c}$ and $|\bar{3}3\rangle_{c}$, one can also select the $|11\rangle_{c}$ and $|88\rangle_{c}$ representations when constructing the tetraquark wave functions. The two sets of color configurations are equivalent to each other. The $|6\bar{6}\rangle_{c}$ and $|\bar{3}3\rangle_{c}$ configurations can be expressed by $|11\rangle_{c}$ and $|88\rangle_{c}$ through the Fierz transformation~\cite{Wang:2019rdo}. Namely, one can extract the $|11\rangle_{c}$ and $|88\rangle_{c}$ components in a physical states expressed with the $|6\bar{6}\rangle_{c}$ and $|\bar{3}3\rangle_{c}$ configurations. A large $|11\rangle_{c}$ component may indicate a potentially large coupling for the initial tetraquark coupling to a nearby $S$-wave channel.

Due to different symmetries of wave functions, there are slight differences in the calculation methods between the $cs\bar{c}\bar{s}$ and $cc\bar{s}\bar{s}$ systems.
Taking the physical state $T_{cs\bar{c}\bar{s}\ 2^{++}\left(4389\right)}$ as an example,
it shows that there exist configurations of different colors, but with the same spatial, flavor, and spin wave functions. For convenience, taking the physical state $T_{cs\bar{c}\bar{s}\ 2^{++}\left(4389\right)}$ that only includes two configurations as an example, the wave function for the physical state is given by:
\begin{eqnarray}
T_{cs\bar{c}\bar{s}\ 2^{++}\left(4389\right)} & = & c_{1}\chi_{22}^{11}\left|6\bar{6}\right\rangle_c +c_{2}\chi_{22}^{11}\left|\bar{3}3\right\rangle_c \ ,
\end{eqnarray}
where $c_{1}$ and $c_{2}$ are the elements of the mixing matrix given in Table~\ref{tab:mass:cscs:1S}.
Expressing the wave function in the basis of $\left|11\right\rangle_c /\left|88\right\rangle_c $,
we have:
\begin{eqnarray}
T_{cs\bar{c}\bar{s}\ 2^{++}\left(4389\right)} & = & \left(c_{1}\ \sqrt{\frac{2}{3}}+c_{2}\ \sqrt{\frac{1}{3}}\right)\chi_{22}^{11}\left|11\right\rangle_c \nonumber \\
& + & \left(c_{1}\ \sqrt{\frac{1}{3}}-c_{2}\ \sqrt{\frac{2}{3}}\right)\chi_{22}^{11}\left|88\right\rangle_c \ .
\end{eqnarray}
So the probabilities for $T_{cs\bar{c}\bar{s}\ 2^{++}\left(4389\right)}$ staying in different color configurations can be extracted by
\begin{eqnarray}\label{ABAB-mixing}
\left|11\right\rangle_c & : & \left|\sqrt{\frac{2}{3}}c_{1}+\sqrt{\frac{1}{3}}c_{2}\right|^{2},\nonumber\\
\left|88\right\rangle_c & : & \left|\sqrt{\frac{1}{3}}c_{1}-\sqrt{\frac{2}{3}}c_{2}\right|^{2}.
\end{eqnarray}
It is interesting to note that for possible mixing elements $c_{1}$ and $c_{2}$, it might happen that a physical state of $cs\bar{c}\bar{s}$ can be dominated by either the $|11\rangle_{c}$ or $|88\rangle_{c}$ configurations. It implies that some tetraquark states may strongly couple to two color-singlet hadrons if the kinematics and dynamics allow.

Carrying on a similar analysis of the $cc\bar{s}\bar{s}$ system and taking the physical state $T_{cc\bar{s}\bar{s}\ 0^{++}\left(4386\right)}$ as an example, the wave function for the physical state is given by $T_{cc\bar{s}\bar{s}\ 0^{++}\left(4386\right)}=c_{1}\ \chi_{00}^{00}\left|6\bar{6}\right\rangle_c +c_{2}\ \chi_{00}^{11}\left|\bar{3}3\right\rangle_c $. It can also be expressed as:
\begin{eqnarray}\label{T-4386}
T_{cc\bar{s}\bar{s}\ 0^{++}\left(4386\right)} & = & \left(c_{1}\ \chi_{00}^{00}\sqrt{\frac{2}{3}}+c_{2}\ \chi_{00}^{11}\sqrt{\frac{1}{3}}\right)\left|11\right\rangle_c \nonumber \\
& + & \left(c_{1}\ \chi_{00}^{00}\sqrt{\frac{1}{3}}-c_{2}\ \chi_{00}^{11}\sqrt{\frac{2}{3}}\right)\left|88\right\rangle_c \ ,
\end{eqnarray}
where $c_{1}$ and $c_{2}$ are the elements of the mixing matrix given in Table~\ref{tab:mass:ccss:1S}.
Note that the spin configurations of $\chi_{00}^{00}$ and $\chi_{00}^{11}$
are linearly independent to each other. They are part of the total wave function for the $|11\rangle_{c}$ and $|88\rangle_{c}$ configurations. Namely, to extract the probability amplitudes for the $|11\rangle_{c}$ and
$|88\rangle_{c}$ components one needs to calculate the spin transition matrix elements first. By taking the orthogonality between $\chi_{00}^{00}$ and $\chi_{00}^{11}$ in Eq. (\ref{T-4386}) we can extract the probabilities for $T_{cc\bar{s}\bar{s}\ 0^{++}\left(4386\right)}$ in the color configurations $|11\rangle_{c}$ and $|88\rangle_{c}$ as follows, respectively:
\begin{eqnarray}\label{AABB-mixing}
\left|11\right\rangle_c & : & \frac{2}{3}c_{1}^{2}+\frac{1}{3}c_{2}^{2},\nonumber\\
\left|88\right\rangle_c & : & \frac{1}{3}c_{1}^{2}+\frac{2}{3}c_{2}^{2}.
\end{eqnarray}
It is interesting to note that for any possible values for the mixing elements $c_{1}$ and $c_{2}$, the color configurations $|11\rangle_{c}$ and $|88\rangle_{c}$ in a physical state can be compatible. This may indicate that an overall color-singlet tetraquark (i.e. $|88\rangle_{c}$ configuration) should always play a role in the $T_{cc\bar{s}\bar{s}}$ system.

Using the wave function obtained by solving the Schr\"{o}dinger equation, the components of different color configurations in the physical $cs\bar{c}\bar{s}$ and $cc\bar{s}\bar{s}$ states can be extracted which are listed in Tables~\ref{tab:squareradius:ABAB} and~\ref{tab:squareradius:AABB}, respectively. It shows that for the $cs\bar{c}\bar{s}$ system, some physical states may have a large $\left|11\right\rangle_c $ component, while some have a large $\left|88\right\rangle_c $. In contrast, for the $cc\bar{s}\bar{s}$ system $\left|11\right\rangle_c $ and $\left|88\right\rangle $ occupy a similar proportion. These numerical values confirms our expectations based on the mixing patterns presented in Eqs. (\ref{ABAB-mixing}) and (\ref{AABB-mixing}). We also note that among those mixing states with the same quantum numbers for the $cs\bar{c}\bar{s}$ system, the higher mass states seem to have a lower component of $|11\rangle_c$. Phenomenologically, the higher mass state in the mixing multiplet is more likely to be an overall color-singlet tetraquark.

\subsection{Root mean square radii}

To investigate the inner structure of the tetraquark, we also calculate the root mean square radius between any two particles. The $\langle\boldsymbol{r}_{12}^{2}\rangle$ is defined as follows:
\begin{equation}
\langle\boldsymbol{r}_{12}^{2}\rangle=\int r_{12}^{2}\left|\Psi\right|^{2}d\boldsymbol{r}_{12}d\boldsymbol{r}_{34}d\boldsymbol{r}_{12-34}.
\end{equation}
Our results are also listed in Tables~\ref{tab:squareradius:ABAB} and~\ref{tab:squareradius:AABB}. The $\langle\boldsymbol{r}_{ij}^{2}\rangle^{\frac{1}{2}}$ represents the average distance between a quark (an antiquark) and an antiquark (a quark). It can be found from the Table~\ref{tab:squareradius:ABAB} and~\ref{tab:squareradius:AABB} that the distances between any two particles share the same size, around 0.6 fm, which is mainly determined by the constituent quark mass.
Our results of $\langle\boldsymbol{r}_{ij}^{2}\rangle^{\frac{1}{2}}$ are consistent with those in Ref.~\cite{Deng:2019dbg}.
The quantity $\langle\boldsymbol{r}_{12-34}^2\rangle^{\frac{1}{2}}$ stands for the averaged distance between the quark pair ($qq$) and antiquark pair ($\bar{q}\bar{q}$).
The quantity $\langle\boldsymbol{r}_{13-24}^2\rangle^{\frac{1}{2}}$ and $\langle\boldsymbol{r}_{14-22}^2\rangle^{\frac{1}{2}}$ stand for the averaged distances between two quark-antiquark pairs ($q\bar{q}$).
The calculated results of $\langle\boldsymbol{r}_{ij-kl}^2\rangle^{\frac{1}{2}}$ are about 0.5 fm, which are higher than those calculated by Ref.~\cite{Deng:2019dbg}.
The averaged distances between two subclusters are less than the averaged distances between two particles ($\langle\boldsymbol{r}_{ij-kl}^2\rangle^{\frac{1}{2}}<\langle\boldsymbol{r}_{ij}^{2}\rangle^{\frac{1}{2}}$), which indicates that the overlap of the two subclusters is extremely strong. In such a case, interpretations of the internal structures based on the hadronic molecules would be problematic. Or at least, one has to include long-ranged dynamics to obtain a more realistic view of the physical states.

\subsection{Rearrangement decays}

The stability of the tetraquark states can be identified by comparing the mass of tetraquark states and threshold of tetraquark states meson-meson thresholds $T_{M_{1}M_{2}}=M_{1}\left(c\bar{s}\right)+M_{2}\left(\bar{c}s\right)$ and $T_{M'_{1}M'_{2}}=M_{1}\left(c\bar{c}\right)+M_{2}\left(s\bar{s}\right)$. Because of the contributions of kinetic energy $\langle T\rangle$ and linear confinement potential $\langle V^{Lin}\rangle$, the mass of all of these states are higher than the corresponding threshold $T_{M_1M_2}$. Tetraquark states with $E>T_{M_{1}M_{2}}$ and $E>T_{M'_{1}M'_{2}}$ are unstable and can decay into two color singlet mesons through the quark rearrangement. It can be found that all the states of $cs\bar{c}\bar{s}$ lie above the thresholds of $\eta_{c}\eta$ and $D_{s}\bar{D_{s}}$. However, it is interesting to find that the state $T_{cs\bar{c}\bar{s}\ 0^{++}}\left(4174\right)$ lies below the thresholds of $D_{s}^{*}\bar{D_{s}^{*}}$. So this state cannot decay into $D_{s}^{*}\bar{D_{s}^{*}}$ through strong interactions in the NRPQM.
While for the $cc\bar{s}\bar{s}$ system, all the states lie above the thresholds of $D_{s}^{*}D_{s}^{*}$. 
So these states can decay into $D_{s}D_{s}$, $D_{s}D_{s}^{*}$, or $D_{s}^{*}D_{s}^{*}$.

In addition to calculating the mass spectra, the results of the fall-apart decays via the quark rearrangement of the $1S$-wave $cs\bar{c}\bar{s}$ and $cc\bar{s}\bar{s}$ states
are given in Tables~\ref{tab:decay:cscs} and~\ref{tab:decay:ccss}.
It shows that all the states of the $cs\bar{c}\bar{s}$ and $cc\bar{s}\bar{s}$ systems, have narrow decay widths within the range of $(0,16)$ MeV. There are two main reasons for it. Firstly, since the initial states are in an $S$ wave, the rearrangement decays are mainly driven by the spin-spin interactions. In contrast, the decay amplitude caused by the confinement potential part $-\frac{3}{16}\left(\boldsymbol{\lambda}_{i}\cdot\boldsymbol{\lambda}_{j}\right)\left(b_{ij}r_{ij}-\frac{4}{3}\frac{\alpha_{ij}}{r_{ij}}+c_{ij}\right)$ is negligibly small. Secondly, one notices that the two terms of the decay amplitude, $\left\langle B_{13}C_{24}|V_{12}+V_{34}|A\right\rangle $ and $\left\langle B_{13}C_{24}|V_{14}+V_{23}|A\right\rangle $ (or $\left\langle B_{14}C_{23}|V_{12}+V_{34}|A\right\rangle $ and $\left\langle B_{14}C_{23}|V_{13}+V_{24}|A\right\rangle $), almost completely cancel out each other.



\begin{table*}
\centering
\caption{The predicted decay widths $\Gamma$ (MeV) of the rearrangement decay processes of the ground $cs\bar{c}\bar{s}$ system. }\centering
\label{tab:decay:cscs}\centering
\tabcolsep=0.21 cm
\begin{tabular}{cccccccccr}
\hline
\hline
State & $\Gamma_{T\rightarrow\eta_{c}\eta}$ & $\Gamma_{T\rightarrow J/\psi\phi}$ & $\Gamma_{T\rightarrow J/\psi\eta}$ & $\Gamma_{T\rightarrow\eta_{c}\phi}$ & $\Gamma_{T\rightarrow D_{s}\bar{D_{s}}}$ & $\Gamma_{T\rightarrow D_{s}^{*}\bar{D_{s}^{*}}}$ & $\Gamma_{T\rightarrow\frac{1}{\sqrt{2}}\left(D_{s}^{*}\bar{D_{s}}-D_{s}\bar{D_{s}^{*}}\right)}$ & $\Gamma_{T\rightarrow\frac{1}{\sqrt{2}}\left(D_{s}^{*}\bar{D_{s}}+D_{s}\bar{D_{s}^{*}}\right)}$ & $\Gamma_{sum}$ \tabularnewline
\hline
$T_{0^{++}}\left(4174\right)$ & \textcolor{blue}{2.0} & 0.07 & \ding{56} & \ding{56} & \textcolor{red}{15} & \ding{56} & \ding{56} & \ding{56} & 18
\tabularnewline
$T_{0^{++}}\left(4356\right)$ & 1.09 & \textcolor{red}{3.4} & \ding{56} & \ding{56} & 0.07 & 0.16 & \ding{56} & \ding{56} & 4.7
\tabularnewline
$T_{0^{++}}\left(4424\right)$ & \textcolor{blue}{4.8} & 0.85 & \ding{56} & \ding{56} & 0.00 & \textcolor{blue}{8.6} & \ding{56} & \ding{56} & 14
\tabularnewline
$T_{0^{++}}\left(4526\right)$ & 0.19 & \textcolor{red}{7.0} & \ding{56} & \ding{56} & 1.3 & 0.03 & \ding{56} & \ding{56} & 8.5
\tabularnewline
$T_{1^{++}}\left(4382\right)$ & \textcolor{red}{\ding{56}} & 0.72 & \ding{56} & \ding{56} & \textcolor{red}{\ding{56}} & \ding{56} & 1.0 & \ding{56} & 1.7
\tabularnewline
$T_{1^{++}}\left(4448\right)$ & \textcolor{red}{\ding{56}} & 0.61 & \ding{56} & \ding{56} & \textcolor{red}{\ding{56}} & \ding{56} & 0.42 & \ding{56} & 1.0
\tabularnewline
$T_{2^{++}}\left(4389\right)$ & \ding{56} & \textcolor{red}{0.80} & \ding{56} & \ding{56} & \ding{56} & \textcolor{blue}{6.0} & \ding{56} & \ding{56} & 6.8
\tabularnewline
$T_{2^{++}}\left(4528\right)$ & \ding{56} & 0.07 & \ding{56} & \ding{56} & \ding{56} & 0.06 & \ding{56} & \ding{56} & 0.13
\tabularnewline
$T_{1^{+-}}\left(4239\right)$ & \ding{56} & \ding{56} & 0.91 & 0.02 & \ding{56} & \textcolor{blue}{2.9} & \ding{56} & \textcolor{blue}{12} & 16
\tabularnewline
$T_{1^{+-}}\left(4322\right)$ & \ding{56} & \ding{56} & 0.03 & 1.18 & \ding{56} & 1.6 & \ding{56} & \textcolor{blue}{4.6} & 7.4
\tabularnewline
$T_{1^{+-}}\left(4492\right)$ & \ding{56} & \ding{56} & 1.1 & 0.52 & \ding{56} & 0.24 & \ding{56} & 1.4 & 3.3
\tabularnewline
$T_{1^{+-}}\left(4529\right)$ & \ding{56} & \ding{56} & 1.1 & 1.6 & \ding{56} & 0.53 & \ding{56} & 0.14 & 3.4
\tabularnewline
\hline
\hline
\end{tabular}
\end{table*}

\begin{table*}
\centering
\caption{Candidates for the $cs\bar{c}\bar{s}$.}
\label{tab:particle:ccX}
\tabcolsep=0.3 cm
\begin{tabular}{cclcccl}
\hline
\hline
State & & Mass(MeV) & Width(MeV) & Decay & $J^{P\left(C\right)}$ & Ref. \tabularnewline
\hline
$X\left(3960\right)$ & $3960<T_{0^{++}}\left(4174,4526\right)$ & $3956\pm5\pm10$ & $43\pm13\pm8$ & $B^{+}\to\textcolor{red}{D_{s}^{-}D_{s}^{+}}K^{+}$ & $0^{++}$ &~\cite{LHCb:2022vsv} \tabularnewline
$X_{0}\left(4140\right)$ & $\Gamma_{T_{0^{++}}\left(4174\right)\to D_{s}\bar{D_{s}}}=15.44$ & $4133\pm6\pm6$ & $67\pm17\pm7$ & $B^{+}\to\textcolor{red}{D_{s}^{-}D_{s}^{+}}K^{+}$ & $0^{++}$ &~\cite{LHCb:2022vsv} \tabularnewline
$X\left(4500\right)$ & $\Gamma_{T_{0^{++}}\left(4526\right)\rightarrow J/\psi\ \phi\left(1020\right)}=7.01$ & $4506\,_{-\,19}^{+\,16}$ & $92_{-21}^{+30}$ & $B^{+}\to\textcolor{blue}{J/\psi\phi}K^{+}$ & $0^{++}$ &~\cite{LHCb:2016axx} \tabularnewline
$X\left(4700\right)$ & $4700>T_{0^{++}}\left(4174,4526\right)$ & $4704\,_{-\,26}^{+\,17}$ & $120_{-45}^{+52}$ & $B^{+}\to\textcolor{blue}{J/\psi\phi}K^{+}$ & $0^{++}$ &~\cite{LHCb:2016axx} \tabularnewline
$X\left(4350\right)$ & $\Gamma_{T_{0^{++}}\left(4356\right)\rightarrow J/\psi\ \phi\left(1020\right)}=3.38$, & $4350.6_{-5.1}^{+4.6}\pm0.7$ & $13.3_{-9}^{+18}\pm4$ & $\gamma\gamma\to\textcolor{blue}{J/\psi\phi}$ & $0/2^{++}$ &~\cite{Belle:2009rkh} \tabularnewline
& $\Gamma_{T_{2^{++}}\left(4389\right)\rightarrow J/\psi\ \phi\left(1020\right)}=0.80$ & & & & & \tabularnewline
$X\left(4140\right)$ & $4140<T_{1^{++}}\left(4382,4448\right)$ & $4143.0\pm2.9\pm1.2$ & $11.7_{-5.0}^{+8.3}\pm3.7$ & $B^{+}\to\textcolor{blue}{J/\psi\phi}K^{+}$ & $1^{++}$ &~\cite{CDF:2009jgo,CMS:2013jru,D0:2013jvp,LHCb:2016axx} \tabularnewline
$X\left(4274\right)$ & $4274<T_{1^{++}}\left(4382,4448\right)$ & $4274_{-6.7}^{+8.4}\pm1.9$ & $\Gamma=32.3_{-15.3}^{+21.9}\pm7.6$ & $B\to\textcolor{blue}{J/\psi\phi}K$ & $1^{++}$ &~\cite{CDF:2011pep,LHCb:2016axx} \tabularnewline
$X\left(4685\right)$ & $4685>T_{1^{++}}\left(4382,4448\right)$ & $4684\,_{-\,17}^{+\,15}$ & $126\pm15\,_{-41}^{+37}$ & $B^{+}\to\textcolor{blue}{J/\psi\phi}K^{+}$ & $1^{+}$ &~\cite{LHCb:2021uow} \tabularnewline
\hline
\hline
\end{tabular}
\end{table*}

\subsection{Discussions}

Based on the tetraquark picture, the $1S$ states for the $T_{cs\bar{c}\bar{s}}$ and $T_{cs\bar{c}\bar{s}}$ have relatively simple spectra in the NRPQM. In this Subsection we discuss the experimental observations and their possible assignments to the calculated spectra.

\subsubsection{$T_{cs\bar{c}\bar{s}}$}

\begin{itemize}
\item $J^{PC}=0^{++}$

In the $J^{PC}=0^{++}$ sector, there are four $1S$ states for the $T_{cs\bar{c}\bar{s}}$ system. Including configuration mixing effects, their physical masses are predicted to be in the range of $\left[4174,4526\right]$ MeV (see Table~\ref{tab:mass:cscs:1S}), which are much larger than the mass threshold of $D_{s}\bar{D_{s}}$. Thus, they may easily decay into $\eta_{c}\eta$, $J/\psi\phi$, or $D_{s}\bar{D_{s}}$ through the quark rearrangements. For the ones above the $D_{s}^{*}\bar{D_{s}^{*}}$ threshold, they can also decay into $D_{s}^{*}\bar{D_{s}^{*}}$ via an $S$ wave \footnote{The $D$-wave transition, in principle, can occur via the spin-orbital tensor coupling. However, this transition will be highly suppressed at the order of $(p/\mu)^2$ with $p$ and $m$ denoting the typical momentum and reduced mass of two interacting constituent quarks. For the ground state tetraquarks investigated in this work we have neglected the $D$-wave contributions as a reasonable approximation. }. The calculated partial decay widths are listed in Table~\ref{tab:decay:cscs}.

The predicted masses of $T_{cs\bar{c}\bar{s}\ 0^{++}}$ are much higher than the state $X\left(3960\right)$ observed by LHCb~\cite{LHCb:2022vsv}. Because of this, it is difficult to accommodate the states $X\left(3960\right)$ as one of the $1S$ states of $T_{cs\bar{c}\bar{s}}$ with $J^{PC}=0^{++}$ in the NRPQM.
The investigation in an improved chromomagnetic interaction model~\cite{Guo:2022crh} and an extended recoupling model~\cite{Badalian:2023qyi} indicates that $X\left(3960\right)$ may be interpreted as $0^{++}$ $cs\bar{c}\bar{s}$
tetraquark states. In Ref.~\cite{Li:2023wxm}, the $X\left(3960\right)$ is interpreted as tetraquark $cs\bar{c}\bar{s}$ states by assuming that the $X\left(4140\right)$ is the lower $1^{++}$ $cs\bar{c}\bar{s}$ tetraquark.
Apart from the interpretations which treat $X\left(3960\right)$ as genuine tetraquark states, there are also other explanations proposed in the literature. For instance, in Refs.~\cite{Bayar:2022dqa,Ji:2022uie,Ji:2022vdj,Xin:2022bzt,Mutuk:2022ckn,Chen:2022dad} the authors interpreted it as a hadronic molecule.

The lowest energy of $T_{cs\bar{c}\bar{s}\ 0^{++}}$ is quite close to the mass of the state $X_{0}\left(4140\right)$, and the partial width $\Gamma_{T_{cs\bar{c}\bar{s}\ 0^{++}}\left(4174\right)\to D_{s}\bar{D_{s}}}$ is predicted to be $\sim15$ MeV, which implies a possibility that the main component of the state $X_{0}\left(4140\right)$ may be the state $cs\bar{c}\bar{s}$ with $J^{PC}=0^{++}$.
In Refs.~\cite{Li:2023wxm,Guo:2022crh,Badalian:2023qyi}, the $X_{0}\left(4140\right)$ is interpreted as tetraquark $cs\bar{c}\bar{s}$ states
while the conclusion that the $X_{0}\left(4140\right)$ is a $D_{s}^{+}D_{s}^{-}$ molecule is drawn in Ref.~\cite{Agaev:2023gti}.

In our calculation results, there are two states, $T_{cs\bar{c}\bar{s}\ 0^{++}}\left(4356\right)$ and $T_{cs\bar{c}\bar{s}\ 2^{++}}\left(4389\right)$, close to $4350$ MeV. And the partial widths of $\Gamma_{T_{cs\bar{c}\bar{s}\ 0^{++}}\left(4356\right)\rightarrow J/\psi\phi}$ and $\Gamma_{T_{cs\bar{c}\bar{s}\ 2^{++}}\left(4389\right)\rightarrow J/\psi\phi}$ are predicted to be $3.38$ and $0.80$ MeV, respectively. It implies a possibility that the main component of the state $X\left(4350\right)$ may be the state $T_{cs\bar{c}\bar{s}\ 0^{++}}\left(4356\right)$.
In Refs.~\cite{Li:2023wxm,Yang:2019dxd,Deng:2019dbg,Wu:2016gas}, the $X\left(4350\right)$ was explained as an $cs\bar{c}\bar{s}$ tetraquark state with $J^{PC}=0^{++}$ or $2^{++}$. However, the state cannot be interpreted as a $cs\bar{c}\bar{s}$ tetraquark with either $0^{++}$ and $2^{++}$ in the QCD sum rules~\cite{Chen:2017dpy}. The $X\left(4350\right)$ can also be accommodated by the charmonium spectrum within an unquenched quark model including coupled-channel effects~\cite{Deng:2023mza}.

The highest mass state $T_{cs\bar{c}\bar{s}\ 0^{++}}(4526)$ may be a candidate of $X\left(4500\right)$. The partial width $\Gamma_{T_{cs\bar{c}\bar{s}\ 0^{++}}\left(4526\right)\rightarrow J/\psi\phi}$ is predicted to be $\sim7.$ MeV, 
which is about 10\% of the measured total width $\Gamma_{exp}\simeq92_{-21}^{+30}$ MeV of $X\left(4500\right)$. The predicted masses for the $0^{++}$ states are much lower than that of the state $X\left(4700\right)$. In this way, it is difficult to accommodate the $X\left(4700\right)$ as $1S$ states for $T_{cs\bar{c}\bar{s}}$ with $J^{PC}=0^{++}$ in the NRPQM. However, we cannot rule out the possibility of interpreting the state $X\left(4700\right)$ as excited state for $cs\bar{c}\bar{s}$ system.
And because of their high masses, the $X\left(4500\right)$ and $X\left(4700\right)$ may be interpreted as the orbitally or radially excited tetraquark or molecular states~\cite{Chen:2016oma,Maiani:2016wlq,Lu:2016cwr,Deng:2019dbg,Zhu:2016arf,Wu:2016gas}.
In addition to the $cs\bar{c}\bar{s}$ explanation,
$X\left(4500\right)$ and $X\left(4700\right)$ can also be accommodated by the charmonium
spectrum~\cite{Deng:2023mza,Ortega:2016hde}.

In brief, the $X_{0}\left(4140\right)$, $X\left(4350\right)$, and $X\left(4500\right)$ may be interpreted as the $1S$ states for the $T_{cs\bar{c}\bar{s}}$ system with $J^{PC}=0^{++}$.
The second highest $0^{++}$ state has mass $4424$ MeV. Its partial widths into $\eta_{c}\eta$ and $D_{s}^{*}\bar{D}_{s}^{*}$ channels are predicted to be $\Gamma_{T_{cs\bar{c}\bar{s}\ 0^{++}}\left(4424\right)\to\eta_{c}\eta}\simeq 4.8$ MeV and $\Gamma_{T_{cs\bar{c}\bar{s}\ 0^{++}}\left(4424\right)\to D_{s}^{*}\bar{D_{s}^{*}}}\simeq8.6$ MeV, respectively. Experimental search for $T_{cs\bar{c}\bar{s}\ 0^{++}}\left(4424\right)$ in $\eta_{c}\eta$ or $D_{s}^{*}\bar{D_{s}^{*}}$ is strongly recommended.

\item $J^{PC}=1^{++}$

In the $J^{PC}=1^{++}$ sector, there are two $1S$ states for the $T_{cs\bar{c}\bar{s}}$
system. Including configuration mixing effects, the physical masses are predicted to be in the range of $\left[4382,4448\right]$ MeV (see Table~\ref{tab:mass:cscs:1S}). Although their masses are much larger
than the mass threshold of $J/\psi\phi$, it shows that the two states have rather narrow decay widths within
the range of $\sim 0-1$ MeV. This feature appears in the hidden-charm decays of the $T_{cs\bar{c}\bar{s}}$ system which can be explained by the mismatching of the wave function overlap between the heavy-heavy and light-light clusters.

In our calculation the lowest mass state with $J^{PC}=1^{++}$ is $4382$ MeV, which is much higher than the two observed states $X\left(4140\right)$ and $X\left(4270\right)$. This could be an indication that additional dynamics in addition to the quark potential should be present. Similar to the situation in the charmonium spectrum that the calculated mass of $\chi_{c1}(2P)$ in the potential quark model is much higher than the observed $X(3872)$. A broadly accepted explanation is that the strong coupling between $\chi_{c1}(2P)$ and the $D\bar{D}^*+c.c.$ threshold may provide a source of the short-distance dynamics for the $D\bar{D}^*+c.c.$ interaction, and the unitarization of the $D\bar{D}^*+c.c.$ final state interaction will dynamically generate a pole, and shift the bare pole mass (i.e. the mass of $\chi_{c1}(2P)$) to the physical one near threshold (see e.g. the review of Ref.~\cite{Guo:2017jvc}).

In the studies of QCD sum rules~\cite{Chen:2016oma} and simple color-magnetic interaction models~\cite{Wu:2016gas} the states, $X\left(4140\right)$ and $X\left(4274\right)$, are interpreted as $S$-wave tetraquark states with $1^{++}$.
In contrast, $X\left(4140\right)$ was the only state explained as an $cs\bar{c}\bar{s}$ tetraquark state with $J^{PC}=1^{++}$ in Refs.~\cite{Li:2023wxm,Wang:2021ghk,Maiani:2016wlq,Lu:2016cwr},
while in Refs.~\cite{Liu:2021xje,Deng:2019dbg,Yang:2019dxd}, $X\left(4270\right)$ was the only state explained as an $cs\bar{c}\bar{s}$ tetraquark with $J^{PC}=1^{++}$.
In Refs.~\cite{Deng:2023mza,Ortega:2016hde,Lu:2016cwr}, the $X\left(4274\right)$ was assigned as a good candidate
for the charmonium.

The predicted $1S$ states with $J^{PC}=1^{++}$ in the NRPQM are much lower than that of $X\left(4685\right)$ observed by LHCb~\cite{LHCb:2021uow}. However, we cannot rule out the possibility of $X\left(4685\right)$ and $X\left(4630\right)$ being the excited states for the $cs\bar{c}\bar{s}$ system.

\item $J^{PC}=2^{++}$

In the $J^{PC}=2^{++}$ sector, two $1S$ states are predicted for the $T_{cs\bar{c}\bar{s}}$ system with configuration mixing included. The physical masses are predicted to be in the range of $\left[4389,4528\right]$ MeV (see Table~\ref{tab:mass:cscs:1S}). Their masses are sufficient for decaying into $J/\psi\phi$ and $D_{s}^{*}\bar{D_{s}^{*}}$, and as shown in Table~\ref{tab:decay:cscs} the partial decay widths turn out to be rather small. It is interesting to see that the partial decay width $\Gamma_{T_{cs\bar{c}\bar{s}\ 2^{++}}\left(4389\right)\rightarrow D_{s}^{*}\bar{D_{s}^{*}}}=6.0$ MeV is much larger than $\Gamma_{T_{cs\bar{c}\bar{s}\ 2^{++}}\left(4528\right)\rightarrow D_{s}^{*}\bar{D_{s}^{*}}}=0.06$ MeV, although the mass of $T_{cs\bar{c}\bar{s}\ 2^{++}}\left(4389\right)$ is smaller.

Recall that the observed state $X\left(4350\right)$ in $\gamma\gamma\to J/\psi\phi$~\cite{Belle:2009rkh} may have $J^{PC}=0^{++}$ or $2^{++}$, and its width is $13.3_{-9}^{+18}\pm 4$ MeV. It seems that it may be assigned as either
$T_{cs\bar{c}\bar{s}\ 0^{++}}\left(4356\right)$ or $T_{cs\bar{c}\bar{s}\ 2^{++}}\left(4389\right)$ within the error. As shown in Table~\ref{tab:decay:cscs} that the partial decay width of $\Gamma_{T_{cs\bar{c}\bar{s}\ 2^{++}}\left(4389\right)\to D_{s}^{*}\bar{D_{s}^{*}}}=6.0$ MeV is significantly larger than that of $T_{cs\bar{c}\bar{s}\ 0^{++}}\left(4356\right)$. This feature may be useful for identifying these two states in experiment. At this moment the measured width of $X\left(4350\right)$ still has large uncertainties. We anticipate that an improved measurement of its total width may allow us to understand better its internal structure.

\item $J^{PC}=1^{+-}$

In the $J^{PC}=1^{+-}$ sector, there are four $1S$ mixing states in the range of $\left[4239,4529\right]$ MeV (see Table~\ref{tab:mass:cscs:1S}). Again, we see that their rearrangement decay widths are generally small, although they have sufficiently large phase spaces for their decays into various channels. In Table~\ref{tab:decay:cscs} their partial decay widths into $J/\psi\eta$, $\eta_{c}\phi$, $D_{s}\bar{D_{s}^{*}}/D_{s}^{*}\bar{D_{s}}$, or $D_{s}^{*}\bar{D_{s}^{*}}$ through quark rearrangements are listed. It is interesting to note that the lightest state $T_{cs\bar{c}\bar{s}\ 1^{+-}}\left(4239\right)$ appears to have the largest partial decay widths into $D_{s}^{*}\bar{D}_{s}^{*}$ and $D_s^*\bar{D}_s+c.c.$, i.e.
$\Gamma_{T_{1^{+-}}\left(4239\right)\to D_{s}^{*}\bar{D}_{s}^{*}}\simeq2.9$ MeV
and $\Gamma_{T_{1^{+-}}\left(4239\right)\to D_{s}^{*}\bar{D_{s}}+c.c.}\simeq12$ MeV, respectively. These large partial decays widths indicate large couplings for $T_{cs\bar{c}\bar{s}\ 1^{+-}}\left(4239\right)$ to the $D_{s}^{*}\bar{D}_{s}^{*}$ and $D_s^*\bar{D}_s+c.c.$ channels. Taking into account the closeness of $T_{cs\bar{c}\bar{s}\ 1^{+-}}\left(4239\right)$ to the $D_{s}^{*}\bar{D}_{s}^{*}$ threshold, it may imply that $T_{cs\bar{c}\bar{s}\ 1^{+-}}\left(4239\right)$ could be a good candidate combing both tetraquark and hadronic molecule dynamics. The tetraquark configuration can provide a source of the short-distance dynamics while the unitarized interactions between $D_{s}^{*}\bar{D}_{s}^{*}$ may lead to pole structures near the $D_{s}^{*}\bar{D}_{s}^{*}$ threshold. Experimental search for this state in both $D_{s}^{*}\bar{D}_{s}^{*}$ and $D_s^*\bar{D}_s+c.c.$ channels is strongly recommended.

\end{itemize}

\subsubsection{$T_{cc\bar{s}\bar{s}}$}

\begin{table}
\centering
\caption{The predicted decay widths $\Gamma$ (MeV) of the rearrangement decay processes of the ground $cc\bar{s}\bar{s}$ system. }
\label{tab:decay:ccss}\centering
\tabcolsep=0.25 cm
\begin{tabular}{ccccc}
\hline
\hline
State & $\Gamma_{T\rightarrow D_{s}\bar{D}_{s}}$ & $\Gamma_{T\rightarrow D_{s}^{*}\bar{D}_{s}^{*}}$ & $\Gamma_{T\rightarrow \bar{D}_{s}D_{s}^{*}+c.c.}$ & $\Gamma_{sum}$ \tabularnewline
\hline
$T_{0^{+}}\left(4386\right)$ & 0.56 & 0.05 & \ding{56} & 0.61
\tabularnewline
$T_{0^{+}}\left(4510\right)$ & \textcolor{blue}{2.85} & \textcolor{blue}{3.55} & \ding{56} & 6.40
\tabularnewline
$T_{1^{+}}\left(4449\right)$ & \textcolor{red}{\ding{56}} & \ding{56} & 0.46 & 0.46
\tabularnewline
$T_{2^{+}}\left(4481\right)$ & \ding{56} & \textcolor{blue}{13.85} & \ding{56} & 13.85
\tabularnewline
\hline
\hline
\end{tabular}
\end{table}

\begin{itemize}

\item $J^{P}=0^{+}$

In the $J^{P}=0^{+}$ sector, there are two $1S$ mixing states with $J^{P}=0^{+}$ in the range of $\left[4386,4510\right]$ MeV (see Table~\ref{tab:mass:ccss:1S}). It shows that their masses are much larger than the mass threshold of $D_{s}^{*}D_{s}^{*}$. Thus, they may easily decay into $D_{s}\bar{D}_{s}$ and $D_{s}^{*}\bar{D}_{s}^{*}$ in an $S$ wave. As listed in Table~\ref{tab:decay:ccss}, its partial decay widths for the $D_{s}\bar{D}_{s}$ and $D_{s}^{*}\bar{D}_{s}^{*}$ channels are rather small, which are just 0.56 and 0.05 MeV, respectively. In contrast, $T_{cc\bar{s}\bar{s}\ 2^{++}}\left(4510\right)$ decays into $D_{s}\bar{D}_{s}$ with a width of 2.9 MeV, and a significantly large partial decay width of 3.6 MeV into the $D_{s}^{*}D_{s}^{*}$ channel. This indicates a strong coupling for $T_{cc\bar{s}\bar{s}\ 2^{++}}\left(4510\right)$ to the $D_{s}^{*}\bar{D}_{s}^{*}$ channel. Experimental search for these two states in the $D_{s}\bar{D}_{s}$ and $D_{s}^{*}\bar{D}_{s}^{*}$ channels seems to be promising.

\item $J^{P}=1^{+}$

In the $J^{P}=1^{+}$ sector, there is only one $1S$ state with $J^{P}=1^{+}$ and a mass of $4449$ MeV (see Table~\ref{tab:mass:ccss:1S}). Its mass is much larger than the mass threshold of $\bar{D}_{s}D_{s}^{*}+c.c.$. However, the calculation shows that its partial decay width $\Gamma_{T_{cc\bar{s}\bar{s}\ 1^{+}}\left(4449\right)\rightarrow D_{s}D_{s}^{*}+c.c.}\simeq0.46$ MeV is a rather small value (see Table~\ref{tab:decay:ccss}). While it may be difficult to observe this state in experiment based on the present experimental statistics, a search of its existence may provide a peculiar evidence for the exotic object as a genuine tetraquark.

\item $J^{P}=2^{+}$

In the $J^{P}=2^{+}$ sector, there is only one $1S$ state with $J^{P}=2^{+}$ and a mass of $4481$ MeV (see Table~\ref{tab:mass:ccss:1S}). Its mass is much larger than the mass threshold of $D_{s}^{*}\bar{D}_{s}^{*}$. The calculated partial decay width $\Gamma_{T_{cc\bar{s}\bar{s}\ 2^{+}}\left(4481\right)\to D_{s}^{*}\bar{D}_{s}^{*}}\simeq14$ MeV is listed Table~\ref{tab:decay:ccss}. Our calculation suggests that $T_{cc\bar{s}\bar{s}\ 2^{+}}(4481)$ is also a narrow state. Its sizeable coupling to the $D_{s}^{*}\bar{D}_{s}^{*}$ channel implies that it has a good chance to be observed in the $D_{s}^{*}\bar{D}_{s}^{*}$ decay channel.

\end{itemize}

\section{summary} \label{SUM}

In this work we present a systematic study of the $1S$-wave hidden and double charm-strange tetraquarks $cs\bar{c}\bar{s}$ and $cc\bar{s}\bar{s}$ in the NRPQM.
The explicitly correlated Gaussian method~\cite{Mitroy:2013eom,Varga:1995dm} is adopted to solve the four-body problem numerically, and the mass spectra, color-spin configurations and possible decay modes are obtained. It is worth noting that for such four-body systems, the $1S$-wave spectra have become rich enough for producing many interesting phenomena. We find that although these states are all above their open flavor thresholds their rearrangement decay widths are rather narrow which can be understood by the mismatching of the wave functions between the initial and final states. It implies that tetraquarks of $cs\bar{c}\bar{s}$ and $cc\bar{s}\bar{s}$ may have a better chance to exist as genuine tetraquark states. Meanwhile, by extracting the color-spin configurations of the $cs\bar{c}\bar{s}$ and $cc\bar{s}\bar{s}$ systems we find that for a physical state of $cs\bar{c}\bar{s}$ its color configurations can be dominated by either the $|11\rangle_{c}$ or $|88\rangle_{c}$ ones. It suggests that some hidden charm-strange tetraquark states may strongly couple to two color-singlet hadrons if the kinematics and dynamics allow. In contrast, we find that the color configurations $|11\rangle_{c}$ and $|88\rangle_{c}$ in a double charm-strange $cc\bar{s}\bar{s}$ state can be compatible. It may suggest that an overall color-singlet tetraquark (i.e. $|88\rangle_{c}$ configuration) should always play a role in the $T_{cc\bar{s}\bar{s}}$ states.

We also make an analysis of the calculated spectra taking into account some experimental candidates. For the hidden charm-strange system we find that $X_{0}\left(4140\right)$, $X\left(4350\right)$, and $X\left(4500\right)$ can be interpreted as the $1S$ $T_{cs\bar{c}\bar{s}}$ states with $J^{PC}=0^{++}$. While there are four $J^{PC}=0^{++}$ states are expected, the experimental search for the predicted $T_{cs\bar{c}\bar{s}\ 0^{++}}\left(4424\right)$ in $\eta_{c}\eta$ or $D_{s}^{*}\bar{D_{s}^{*}}$ is strongly recommended. Our predictions for the $J^{PC}=1^{++}$ states
seem to have higher masses than the experimental candidates. This is similar to the situation for $X(3872)$ where the potential quark model predicts a higher mass of the $\chi_{c1}(2P)$ state than the experimental observation. This may provide an analogue in the hidden charm-strange sector. Further detailed investigation of the couplings of these $1^{++}$ states to the $D_s\bar{D}_s^*+c.c.$ and $D_s^*\bar{D}_s^*$ channels will be needed. There still lack of experimental evidences for the $J^{PC}=1^{+-}$ states in the hidden charm-strange sector and all the states in the double charm-strange sector. Our calculations shows that these states mostly turn out to be narrow. Thus, the experimental search for their signals at LHCb in some preferred decay channels are strongly recommended.

\begin{acknowledgments}
This work is supported by the National Natural Science Foundation of China (Grants No.12235018, No.12175065,  No.E411645Z10).  Q.Z. is also supported in part, by the DFG and NSFC funds to the Sino-German CRC 110 ``Symmetries and the Emergence of
Structure in QCD'' (NSFC Grant No. 12070131001, DFG Project-ID 196253076), National Key Basic Research
Program of China under Contract No. 2020YFA0406300, and Strategic Priority Research Program of Chinese
Academy of Sciences (Grant No. XDB34030302).
\end{acknowledgments}

\bibliography{refs}

\end{CJK}
\end{document}